\documentclass[%
 reprint,
showpacs,preprintnumbers,
 amsmath,amssymb,
prb,
superscriptaddress
]{revtex4-2}

\usepackage{graphicx}
\usepackage{dcolumn}
\usepackage{bm}
\usepackage{physics}
\usepackage{comment}
\usepackage{algpseudocode}
\graphicspath{{./figures/}}

\usepackage{color}
\newcommand{\redd}[1]{\textcolor{red}{#1}}
\newcommand{\red}[1]{\textcolor{black}{#1}}

\begin{document}


\title{Enhanced energy deposition and carrier generation in silicon induced by two-color intense femtosecond laser pulses}

\author{Mizuki Tani}
\email{mzktani@atto.t.u-tokyo.ac.jp}
\affiliation{%
 Department of Nuclear Engineering and Management, Graduate School of Engineering, The University of Tokyo,7-3-1 Hongo, Bunkyo-ku, Tokyo 113-8656, Japan
}
\author{Kakeru Sasaki}%
\affiliation{%
 Department of Nuclear Engineering and Management, Graduate School of Engineering, The University of Tokyo,7-3-1 Hongo, Bunkyo-ku, Tokyo 113-8656, Japan
}%

\author{Yasushi Shinohara}
\affiliation{%
 Department of Nuclear Engineering and Management, Graduate School of Engineering, The University of Tokyo,7-3-1 Hongo, Bunkyo-ku, Tokyo 113-8656, Japan
}
\affiliation{%
 Photon Science Center, Graduate School of Engineering, The University of Tokyo, 7-3-1 Hongo, Bunkyo-ku, Tokyo 113-8656, Japan
}
\author{Kenichi L. Ishikawa}
\email[corresponding author:]{ishiken@n.t.u-tokyo.ac.jp}
\affiliation{%
 Department of Nuclear Engineering and Management, Graduate School of Engineering, The University of Tokyo,7-3-1 Hongo, Bunkyo-ku, Tokyo 113-8656, Japan
}%
\affiliation{%
 Photon Science Center, Graduate School of Engineering, The University of Tokyo, 7-3-1 Hongo, Bunkyo-ku, Tokyo 113-8656, Japan
}
\affiliation{%
 Research Institute for Photon Science and Laser Technology\, The University of Tokyo, 7-3-1 Hongo, Bunkyo-ku, Tokyo 113-0033, Japan
}




\date{\today}

\begin{abstract}
We theoretically investigate the optical energy absorption of crystalline silicon subject to  dual-color femtosecond laser pulses, using the time-dependent density functional theory (TDDFT). 
We employ the modified Becke-Johnson (mBJ) exchange-correlation potential which reproduces the experimental direct bandgap energy $E_g$. 
We consider situations where the one color is in the ultraviolet (UV) range above $E_g$ and the other in the infrared (IR) range below it.
The energy deposition is examined as a function of mixing ratio $\eta$ of the two colors with the total pulse energy conserved. 
Energy transfer from the laser pulse to the electronic system in silicon is dramatically enhanced by simultaneous dual-color irradiation and maximized at $\eta\sim 0.5$. 
Increased is the number of generated carriers, not the absorbed energy per carrier.
The effect is more efficient for lower IR photon energy, or equivalently, larger vector-potential amplitude.
As the underlying mechanism is identified the interplay between intraband electron motion in the {\it valence} band (before excitation) driven by the IR component and resonant valence-to-conduction interband excitation (carrier injection) induced by the UV component. The former increases excitable electrons which pass through the $k$ points of resonant transitions. The effect of different multiphoton absorption paths or intraband motion of carriers generated in the conduction band play a minor role. 
\end{abstract}

\pacs{Valid PACS appear here}
\maketitle


\section{\label{sec:introduction}INTRODUCTION}
The ultrashort laser ablation of semiconductor and dielectric materials \cite{Tiinnermann,Ilday,Mourou,Mazur} has been attracting increasing attention due to both broad scientific interest and industrial application \cite{Ostrikov2016-sf,Sugioka2014-rb}.
Its advantages include high efficiency and quality \cite{Ilday,Sugioka2021-fh} thanks to the suppression of heat-affected-zone. 
The laser ablation is initiated by transfer of optical energy to electrons.
\red{Then, irreversible damage is left on the material surface when the energy is subsequently transferred to the lattice \cite{Mazur}, the carrier density reaches the critical density where the plasma frequency is identical to the laser frequency \cite{Thorstensen2012-ji}, or the inter-atomic forces are strongly modified owing to massive carrier creation \cite{Rousse2001-ws,Sundaram2002-gc,Shcheblanov2016-eu}.}
Thus, the fundamental understanding of energy transfer from laser pulses to electrons is critical to further improve the efficiency of laser micromachining \cite{Wang2011-pm,Tani2021-Se}.

There have been studies reporting that the use of synthesized dual-color laser pulse pairs enables highly efficient laser ablation of transparent materials, compared to single-color irradiation \cite{Sugioka1993-hg,Zhang1997-of,Zhang2000-ev,Obata2001-ey,Zoppel2005-En,Zoppel2007-Tw,Yu2013-aw,Yu2014-Fa,Yang2015-Fe,Gedvilas2017-Mu}.
In Ref.~\cite{Sugioka1993-hg,Zhang1997-of}, the high efficiency observed under dual-color nanosecond ultraviolet (UV) lasers is attributed to the excited-state absorption (ESA) mechanism where the shorter wavelength component breaks the Si-O covalent bond of fused silica, which increases the absorption efficiency of the longer wavelength component.
References \cite{Zhang2000-ev,Obata2001-ey} have reported that the shorter wavelength laser excites valence electrons into defect or impurity energy levels and, then, that the longer wavelength laser promotes them from the localized levels to conduction states, even if the former alone cannot directly excite valence electrons into the conduction band.
References \cite{Zoppel2005-En,Zoppel2007-Tw} have shown that the combination of picosecond or nanosecond infrared (IR) laser and its 2nd/3rd harmonics also improves ablation efficiency of silicon and 3C-SiC and attributed it to a mechanism where the UV laser excites valence electrons into the conduction band, which are then heated by the IR pulse.
References \cite{Yu2013-aw,Yu2014-Fa} have shown by experiments and analyses based on rate equations that a pair of femtosecond IR laser and its third harmonic UV can reduce ablation threshold of fused silica thanks to a defect state that can be reached by a single UV photon absorption.
It has also been reported that the simultaneous use of femtosecond IR laser and its second harmonic UV enables efficient laser ablation of polymethylmethacrylate \cite{Yang2015-Fe}, since the synthesized electric field has a larger instantaneous value than in single color cases. 
In Ref~.\cite{Gedvilas2017-Mu}, it is argued based on experiments and phenomenological modeling that the modification of sapphire under femtosecond IR laser and its third harmonics is enhanced by the contributions of various multiphoton absorption paths including not only direct valence-conduction but also defect-level-mediated ones.

Photoionization in dielectrics under an intense laser pulse has been theoretically modeled by various methods including the Keldysh theory \cite{Keldysh1965,Shcheblanov2017-em,Otobe2019-cv}, rate equation\cite{Bloembergen1974-hi}, time-dependent schr\"odinger equation within the independent electron approximation (IEA)\cite{Lagomarsino2016-by,PhysRevA.95.043416}, density matrix method\cite{Lindberg1988-pr,DM,PhysRevB.102.041125,HiroriAPL2019,Yue2020-rm}, and time-dependent density functional theory (TDDFT) \cite{Otobe2008-yz,Wachter2014-pg,Yamada2019-ls,Yamada2020-qt,Hui2021-cy}. By modeling fused silica as a two-level system, 
\citet{Martinez2021-Mo} have recently claimed that increase in energy absorption under simultaneous femtosecond IR and UV pulse irradiation is due to the interplay of various multiphoton ionization paths. 
This argument has been supported by \citet{Duchateau2022-ba} also for a band-dispersion system, $\alpha$-quartz. 
\red{Although electron dynamics and high-harmonic generation has been extensively discussed in terms of the coupling between intraband and interband transitions for single-color cases \cite{Golde2008-yi,Ghimire2010-sg,McDonald2015,Wismer2016-hi,PhysRevA.95.043416,DM,Wang2018-cd,Sato2018-Ro,Schlaepfer2018-At,HiroriAPL2019,Song2020-hk,PhysRevB.102.041125}, the role of the coupling under simultaneous two-color irradiation has been little studied.}
Thus, while increase in energy absorption and ablation efficiency by dual-color pulse irradiation appears to be quite general observation, a consensus has not been reached on its mechanisms, which possibly depend on materials and irradiation conditions.

In the present study, we investigate the energy absorption by bulk silicon under simultaneous dual-color (UV and IR) femtosecond laser fields, using numerical simulations based on the time-dependent density functional theory (TDDFT). 
TDDFT is an \textit{ab initio} framework successfully applied to  ultrafast carrier dynamics under intense laser-matter interaction \cite{Kozak2020-Ob,Tani2021-Se,Miyamoto2021-Di,Otobe2016-Hi,Nicolas2017-Im,Nicolas2018-Ul,Nicolas2018-At,Floss2019-In,Yamada2021-De}.
We employ the SALMON code \cite{salmon} and examine the dependence of energy absorption on mixing ratio $\eta$ of of the two color components with the total intensity (or equivalently, fluence and energy) conserved.
Our calculations show that the absorbed energy is significantly enhanced by dual-color irradiation and maximized at $\eta\sim 0.5$.
Our analyses reveal that the intraband motion of the electrons driven in the valence band by the IR field has a substantial role in increasing valence-to-conduction interband transition induced by the UV field. 
%
These observations indicate that the strong-field electron excitation dynamics can be controlled by nonlinear coupling of a long-wavelength-driven intraband motion and a short-wavelength-driven interband transition.

This article is organized as follows. Section \ref{sec:TDDFT} describes our simulation methods. 
We briefly review TDDFT and describe how to evaluate absorbed energy.
Section \ref{sub:GS} shows the calculated ground state and linear response properties of Si.
In Sec.~\ref{sec:results} we present and analyze our numerical results.
Conclusions are given in Sec.~\ref{sec:conclusions}. 

\section{\label{sec:TDDFT}Time-dependent density functional theory}
The time propagation of an $N_e$-electron system under an optical field is calculated by solving the time-dependent Kohn-Sham (TDKS) equations for the Kohn-Sham orbitals $\{\phi_i(\mathbf{r},t)\}$ \cite{TDDFT},
\begin{equation}
    \mathrm{i}\hbar \frac{\partial}{\partial t}\phi_i(\mathbf{r},t) = h_{\mathrm{KS}}[n_{\mathrm{e}}(\mathbf{r},t)]\phi_i(\mathbf{r},t), \label{TDKS}
\end{equation}
where,
\begin{equation}
    h_{\mathrm{KS}}[n_{\mathrm{e}}(\mathbf{r},t)] = \red{\frac{1}{2m}\left[ \mathbf{p}+e\mathbf{A}(t) \right]^2} + V_{\mathrm{eff}}[n_{\mathrm{e}}(\mathbf{r},t)],
\end{equation}
denotes the \red{velocity gauge} Kohn-Sham Hamiltonian within the electric dipole approximation, $\mathbf{p}$ the canonical momentum, $e$ the elementary charge, $\mathbf{A}(t)$ the vector potential, $V_{\rm eff}$ the effective potential (see below), $m$ the electron mass.
The time-dependent electron density $n_e({\bf r},t)$ is given by,
\begin{equation}
    n_{\mathrm{e}}(\mathbf{r},t) = 2\sum_{i\in \mathrm{occ}} |\phi_i(\mathbf{r},t)|^2.
\end{equation}
The effective potential $V_{\mathrm{eff}}$, 
\begin{equation}
    V_{\mathrm{eff}}[n_{\mathrm{e}}(\mathbf{r},t)] = V_{\mathrm{ion}}(\mathbf{r}) + V_{\mathrm{\mathrm{H}}}[n_{\mathrm{e}}(\mathbf{r},t)] + V_{\mathrm{xc}}[n_{\mathrm{e}}(\mathbf{r},t)], \label{Veff}
\end{equation}
consists of the electron-ion potential $V_{\mathrm{ion}}$ consists of the norm-conserving pseudopotential \cite{FHI}, the Hartree potential $V_{\mathrm{H}}$, and the exchange-correlation potential $V_{xc}$. 
We employ the modified Becke-Johnson (mBJ) potential \cite{TBmBJ,Sato2015-No} for $V_{xc}$.

We consider a two-color pulse whose vector potential $\mathbf{A}(t)$ is described by,
\begin{align}
    \mathbf{A}(t) &= \mathbf{A}_1(t)+\mathbf{A}_2(t), \label{At} \\
    \mathbf{A}_1(t) &=  - \mathbf{a}_1\cos^2{\left[ \frac{\pi}{T}\left( t-\frac{T}{2}-t_{\mathrm{delay}} \right) \right]} \notag \\
    &\qquad\qquad\quad \times\sin{\left[ \omega_1\left( t-\frac{T}{2}-t_{\mathrm{delay}} \right) \right]} \notag \\
    &\qquad\qquad\qquad\qquad\quad (t_{\mathrm{delay}}\le t\le t_{\mathrm{delay}}+T), \\
    \mathbf{A}_2(t) &= - \mathbf{a}_2\cos^2{\left[ \frac{\pi}{T}\left( t-\frac{T}{2} \right) \right]}\sin{\left[ \omega_2\left( t-\frac{T}{2} \right) \right]} \notag \\
    &\qquad\qquad\qquad\qquad\qquad\qquad\qquad (0\le t\le T),
\end{align}
where $\mathbf{a}_1,\mathbf{a}_2$ denote the amplitude and polarization vectors of each color, $T$ the foot-to-foot pulse width (corresponding to $0.36T$ in FWHM), $\omega_1, \omega_2$ the central frequency of each color, and $t_{\mathrm{delay}}$ the time delay between the two colors. $\mathbf{A}_1,\mathbf{A}_2$ are zero vectors on outside the time domain of definition ($t<t_{\mathrm{delay}}$ or $t>T+t_{\mathrm{delay}}$ for $\mathbf{A}_1$, $t<0$ or $t>T$ for $\mathbf{A}_2$). We set $t_{\mathrm{delay}}=0$ unless explicitly stated.

We evaluate the absorbed energy as work done by the electric field $\mathbf{E}(t)=-\dot{\mathbf{A}}(t)$,
\begin{equation}
    W=-e\int_0^T \dd \tau \mathbf{J}(\tau) \cdot \mathbf{E}(\tau),
\end{equation}
with the current density,
\begin{equation}
    \begin{split}
        \mathbf{J}(t) = \mathrm{i}\frac{2}{\red{\hbar}\Omega} \int_{\Omega} d \mathbf{r} \sum_{i} \phi_{i}^{*}(\mathbf{r}, t) \left[h_{\mathrm{KS}},\mathbf{r}\right] \phi_{i}(\mathbf{r}, t), \label{Jt}
    \end{split}
\end{equation}
where $\Omega$ is the volume of the simulation box.

Using the open source package SALMON\cite{salmon}, our simulation is conducted on a crystalline silicon primitive cell \redd{(see Appendix \ref{sec:appendixC} for the crystal structure and the calculated band structure)}, which is discretized into $16^3$ real-space and $16^3$ $k$-space grids. The lattice constant $a$ is set to 5.468 \AA\,\red{(the experimental value is 5.43 \AA \cite{Massa2009-hp})}. The time step is 0.03 a.u., equivalent to 0.7257 as. \red{These parameters are finer than those used in the previous works \cite{salmon,Yamada2019-ls}. We have also confirmed the convergence of simulation results (see also Appendix \ref{sec:appendixB}). The laser polarization is assumed to be parallel to $\Gamma X$ direction unless specified otherwise.}

\section{\label{sub:GS} Ground state and linear response}

\begin{figure}[tb]
  \centering
  \includegraphics[keepaspectratio,width=\hsize]{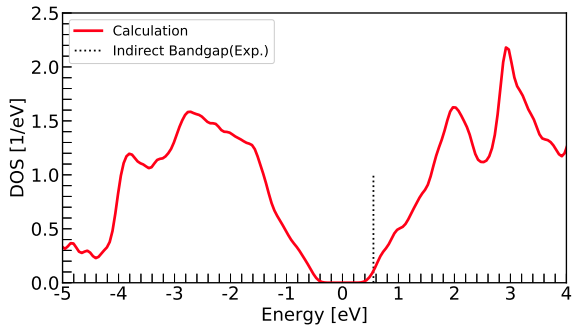}
  \caption[DOS]{The calculated density of states of Si
(red-solid line). The experimental value of the indirect bandgap, 1.1 eV \cite{Sigap}, is indicated by the vertical black-dotted line.}
  \label{DOS}
\end{figure}

The initial ground state of the system is taken as the eigen state of the field-free Hamiltonian $h_{\mathrm{KS}}|_{\mathbf{A}=\mathbf{0}}$.
Figure~\ref{DOS} shows the density of states of the system.
The indirect bandgap is found to be perfectly agreeing with the experimental one, 1.1 eV \cite{Sigap}.

The direct bandgap is estimated by the imaginary part of the dielectric function, obtained by time propagation after an impulsive momentum kick.
This is realized by a step-like vector potential,
\begin{equation}
    \mathbf{A}(t) = \left\{
        \begin{array}{ll}
            -\mathbf{A}_0 & (t \geq 0)\\
            0 & (t < 0)
        \end{array}
    \right. ,
    \label{A0}
\end{equation}
which corresponds to an impulsive electric field,
\begin{equation}
    \mathbf{E}(t)=\mathbf{A}_0\delta(t).
    \label{E0}
\end{equation}
Noting that the electric field has a constant power spectrum over all frequencies, the diagonal component of the optical conductivity is calculated as,
\begin{equation}
    \sigma_{m}(\omega) = \frac{-e\hat{J}_m(\omega)}{A_{0m}}\quad (m = x,y,z),
\end{equation}
where $\hat{J}_m$ ($m,n = x,y,z$) denotes the $m$ component of the temporal Fourier transform of the current density.
Assuming isotropic media, the dielectric function $\varepsilon_{m}(\omega)$ is given by,
\begin{equation}
    \varepsilon_{m}(\omega) = 1+4\pi\mathrm{i}\frac{\sigma_{m}(\omega)}{\omega}.
\end{equation}

\begin{figure}[tb]
  \centering
  \includegraphics[keepaspectratio,width=\hsize]{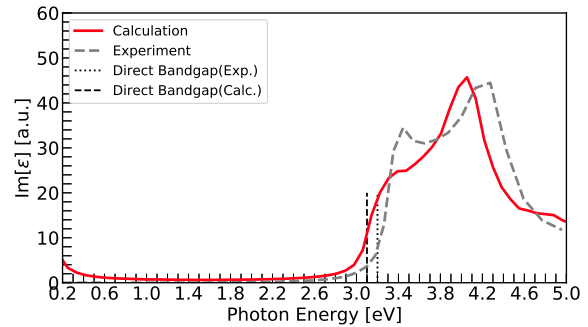}
  \caption[eps]{The calculated imaginary part of the dielectric function of Si (red-solid) and the experimental reference \cite{Sink} (black-dashed). \red{The experimental value of the direct bandgap, 3.1 eV (3.2 eV\cite{Aspnes1983-gs}), is indicated by the vertical black-dashed (black-dotted) line.}}
  \label{eps}
\end{figure}

The imaginary part of thus evaluated dielectric function is consistent with the measured one \cite{Sink} (Fig.~\ref{eps}).
The direct bandgap is found to be 3.1 eV, which quasi-perfectly agrees with the experimental one, 3.2 eV\cite{Aspnes1983-gs}. \red{The direct energy gap, rather than the indirect one (Fig.~\ref{DOS}), is relevant to optical absorption in our calculations.}
Overall, therefore, the prepared ground state reproduces the electronic properties of real crystalline silicon very well.

\section{\label{sec:results}RESULTS AND DISCUSSIONS}

\subsection{\label{subsec:energy}Energy transfer from laser to electrons}

\begin{figure}[tb]
  \centering
  \includegraphics[keepaspectratio,width=\hsize]{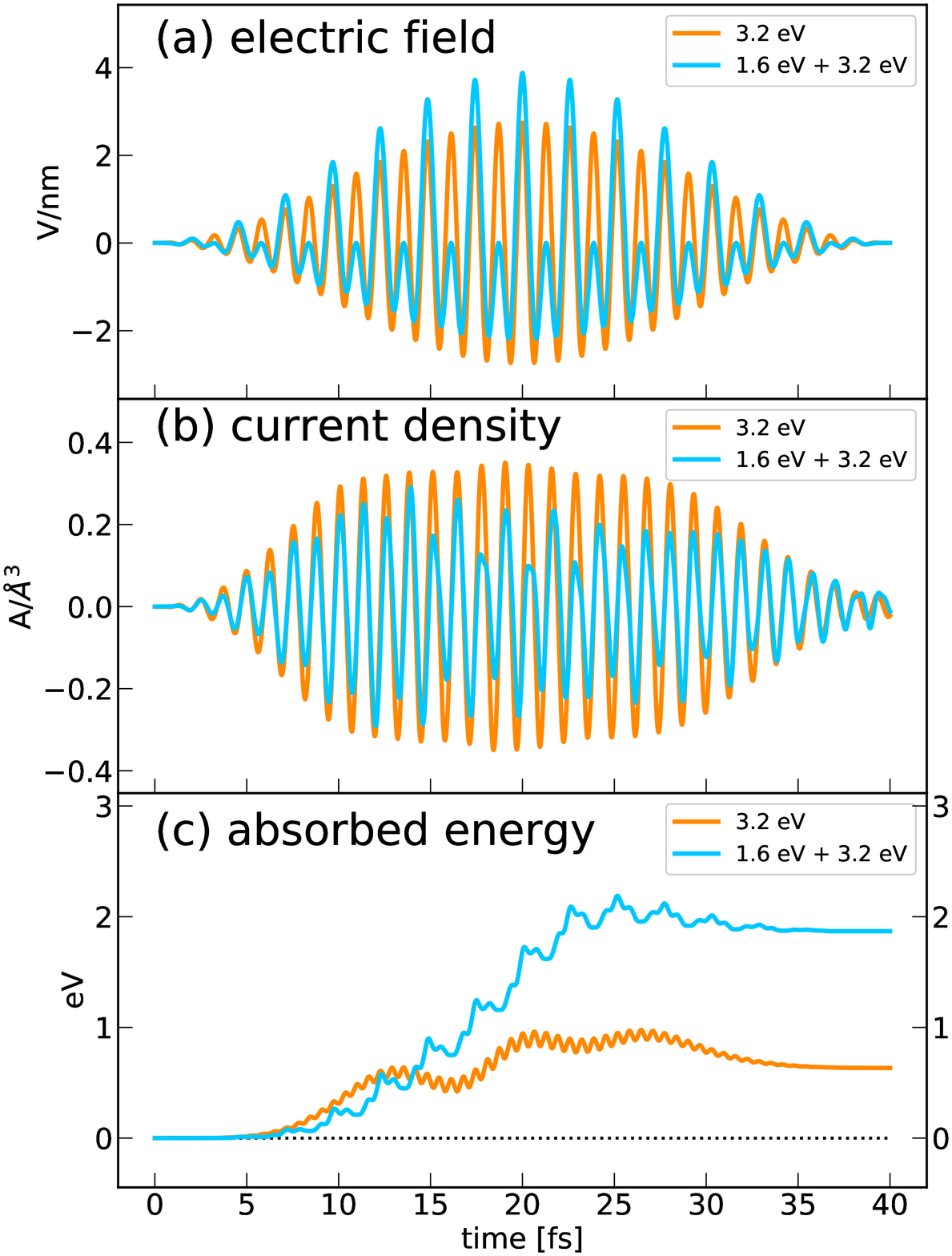}
  \caption[EJW]{Temporal profile of (a) the electric field, (b) the calculated current density, and (c) the calculated absorbed energy for $I_{\rm tot}=10^{12}$ W/$\mathrm{cm}^2$. The orange and cyan solid lines are for the single-color ($\eta=0$) and dual-color ($\eta=\frac{1}{2}$) fields, respectively.}
  \label{EJW}
\end{figure}

Let us consider energy transfer to silicon from superposed two-color fields with $\hbar\omega_1$ and $\hbar\omega_2$ photon energies ($\hbar\omega_1 \le \hbar\omega_2$) and both 14.4 fs FWHM pulse width.
\red{Whereas this pulse width is short compared to those typically used in experiments, computationally demanding TDDFT simulations for such short pulses have been useful to investigate fundamental laser-solid interactions \cite{Yamada2019-ls,Hui2021-cy,Duchateau2022-ba}.}
We examine how the absorbed energy varies with mixing ratio $\eta=I_1/(I_1+I_2)$ with $I_i \,(i=1,2)$ being the peak intensity of color component $i$ while fixing the total intensity $I_{\rm tot}\equiv I_1+I_2$.
Figure~\ref{EJW} displays the temporal profiles of the electric field, the current density, and the absorbed energy for $\hbar\omega_1=1.6$ eV (corresponding to 775 nm wavelength), $\hbar\omega_2=2\hbar\omega_1=3.2$ eV (387.5 nm), and $I_{\rm tot}=10^{12}$ W/$\mathrm{cm}^2$, and compare the results for $\eta=0$ (3.2 eV only) and $\frac{1}{2}$ (1.6 eV and 3.2 eV).

\begin{figure}[tb]
  \centering
  \includegraphics[keepaspectratio,width=\hsize]{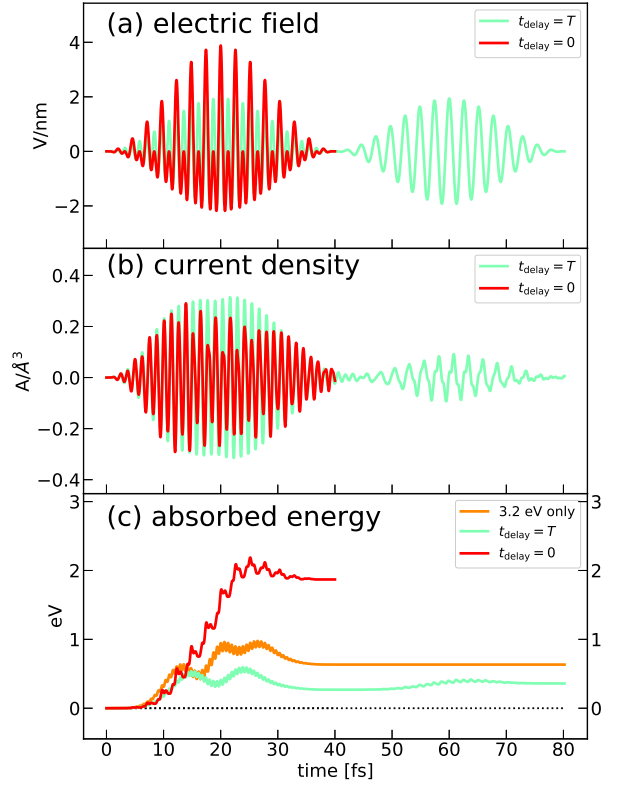}
  \caption[EJW2]{Temporal profile of (a) the electric field, (b) the calculated current density, and (c) the calculated absorbed energy  for $I_{\rm tot}=10^{12}$ W/$\mathrm{cm}^2$ and $\eta=\frac{1}{2}$. Comparison of simultaneous ($t_{\mathrm{delay}}=0$, red solid) and delayed ($t_{\mathrm{delay}}=T$, lime solid)  dual-color irradiation (1.6 eV and 3.2 eV). The absorbed energy under a single-color laser field (3.2 eV) is also plotted in panel (c).}
  \label{EJW2}
\end{figure}

The absorbed energy is significantly (ca. three times) higher for the two-color case.
If we naively assumed an incoherent sum of two-photon excitation by $\hbar\omega_1$ and single-photon ionization by $\hbar\omega_2$, the absorbed energy would scale as $\sigma_1\eta^2 + \sigma_2 (1-\eta)$ with $\sigma_1 (\sigma_2)$ being excitation cross section by $\hbar\omega_1 (\hbar\omega_2)$, from which we would expect decrease, rather than increase, in absorbed energy by two-color mixing.
Indeed, if the two pulses are separated in time, energy absorption is significantly reduced and even smaller than in the single-color case (Fig.~\ref{EJW2}).
Thus, simultaneous, rather than consecutive, irradiation is essential to the enhancement.

\begin{figure}[tb]
  \centering
  \includegraphics[keepaspectratio,width=\hsize]{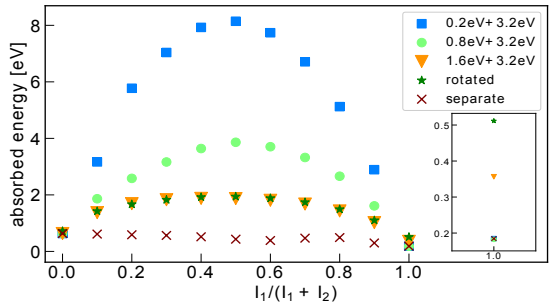}
  \caption[omega1]{Absorbed energy as a function of mixing ratio $\eta$ under $I_{\rm tot}=10^{12}$ W/$\mathrm{cm}^2$ for three different combinations of $\hbar\omega_1$ and $\hbar\omega_2=3.2$ eV indicated in the legend. 
  Green stars: both laser polarization of $\hbar\omega_1=1.6$ eV and $\hbar\omega_2=3.2$ eV are rotated $\pi/4$.
  Brown crosses: delayed irradiation ($t_{\mathrm{delay}}=T$) of $\hbar\omega_1=0.2$ eV and $\hbar\omega_2=3.2$ eV.
  Inset: close-up of the results for $\eta=1$.
  }
  \label{omega1}
\end{figure}

Let us further examine the dependence of the absorbed energy on mixing ratio $\eta$, photon energies $\hbar\omega_{1,2}$, laser polarization orientation, and intensity $I_{\rm tot}$. 
Figure~\ref{omega1} shows the results as a function of $\eta$ for $I_{\rm tot}=10^{12}$ W/$\mathrm{cm}^2$, $\hbar\omega_2=3.2$ eV (387.5 nm wavelength) and three different values of $\hbar\omega_1$.
Note that $\hbar\omega_2$ (the shorter-wavelength component) is larger than the direct band gap.
\begin{figure}[tb]
  \centering
  \includegraphics[keepaspectratio,width=\hsize]{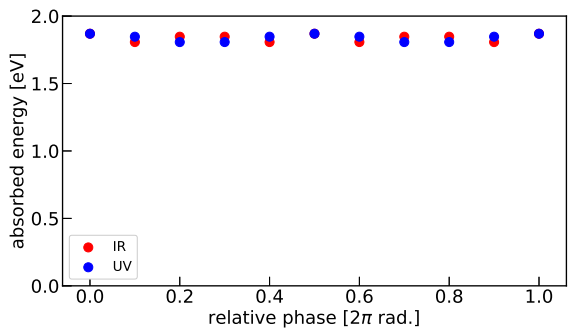}
  \caption[Ephase]{Relative phase dependence of the absorbed energy. Red (blue) circles indicate that the relative phase of the $\hbar \omega_1=1.6$ eV ($\hbar \omega_2=3.2$ eV) field is varied with the $\hbar \omega_2$ ($\hbar \omega_1$) phase fixed.}
  \label{Ephase}
\end{figure}
\begin{figure}[tb]
  \centering
  \includegraphics[keepaspectratio,width=\hsize]{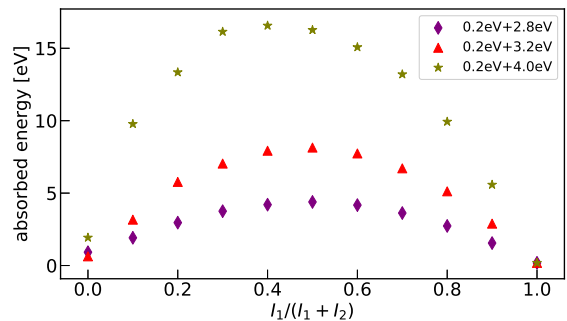}
  \caption[omega2]{Absorbed energy as a function of mixing ratio $\eta$ under $I_{\rm tot}=10^{12}$ W/$\mathrm{cm}^2$ for three different combinations of $\hbar\omega_1=0.2$ eV and $\hbar\omega_2$ indicated in the legend. 
  }
  \label{omega2}
\end{figure}
\red{The green stars in Fig.~\ref{omega1} indicate the transferred energy from the laser field whose polarization is 45° to the $\Gamma X$ direction, they exhibit almost same trend with the original laser polarization. }
Again, combining two colors significantly enhances the energy transfer, 
maximized at $\eta\sim \frac{1}{2}$.
It should be noticed that the enhancement effect is larger with smaller $\hbar\omega_1$, or equivalently, longer wavelength of the first color, in contrast to the fact that, for $\eta=1$, the absorbed energy is smaller with smaller $\hbar\omega_1$.
Since the longer the wavelength, the larger the vector-potential amplitude and, thus, the crystal-momentum shift of electrons for a given laser intensity,
the results in Fig.~\ref{omega1} indicate that $\hbar\omega_1$ (the longer-wavelength component) contributes to energy absorption predominantly through intraband motion, rather than multi-photon interband excitation. \redd{It should be noted that the decomposition into interband and intraband motions is, strictly speaking, gauge dependent. Nevertheless, a physical interpretation relying on a particular gauge is often useful, transparent, and valid, and gives us clear physical insight. Therefore, in this study, we analyze our simulation results by considering the intraband motion as crystal momentum shift $\mathbf{k}(t)=\mathbf{k}_0+\mathbf{A}(t)$, with $\mathbf{k}_0$ being its field-free value.}

The absorbed energy is nearly independent of relative phase between the two color fields (Fig.~\ref{Ephase}). This observation \red{suggests} that tunneling ionization is not dominant in the present situation.
 \red{Indeed, the Keldysh parameter for 1.6 (3.2) eV photon energy and $10^{12}$ W/$\mathrm{cm}^2$ intensity is 5.3 (10), which corresponds to the multi-photon regime.}

In Fig.~\ref{omega2}, which plots the absorbed energy vs.~$\eta$ for $\hbar\omega_1=0.2$ eV (6.2 $\mu {\rm m}$ wavelength) and three different values of $\hbar\omega_2$, 
we can see larger absorption for larger $\hbar \omega_2$ (the shorter-wavelength component), which coincides with the behavior of linear absorption between 2.8 and 4.0 eV (Fig.~\ref{eps}).
This observation suggests that the major role of $\hbar\omega_2$ is to induce interband excitation.

\begin{figure}[tb]
  \centering
  \includegraphics[keepaspectratio,width=\hsize]{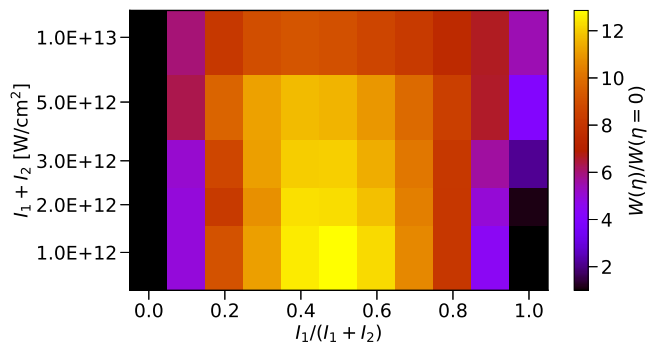}
  \caption[EI]{\red{False-color} representation of the absorbed energy normalized to the case $\eta=0$ [$W(\eta)/W(\eta=0)$] as functions of mixing ratio $\eta$ and total peak intensity ($I_1+I_2=10^{12}-10^{13}$ W/$\mathrm{cm}^2$) for
  $\hbar\omega_1=0.2$ eV (6200 nm wavelength) and $\hbar\omega_2=3.2$ eV (387.5 nm).}
  \label{EI}
\end{figure}

In Fig.~\ref{EI}, we display the absorbed energy normalized to the $\hbar\omega_2$ only case ($\eta=0$) as a function of both $I_{\rm tot}$ and $\eta$
at $\hbar\omega_1=0.2$ eV and $\hbar\omega_2=3.2$ eV.
\red{The lower intensities exhibit higher maximum-to-minimum contrast (ca.~12 for $I_{\rm tot}=10^{12}$ W/$\mathrm{cm}^2$ and 9 for $10^{13}$ W/$\mathrm{cm}^2$) and symmetric profiles, which become asymmetric with the peak position slightly shifting to the left for higher total intensities. Nevertheless, the absorption is enhanced when both wavelengths are approximately equally mixed, within the intensity range investigated here.}

\subsection{\label{subsec:excited}Number of excited electrons and mean absorbed energy}

The absorbed energy is decomposed into the number of generated carriers (excited electrons) and their mean energy absorption. 
Then, is it increase in the former or the latter that accounts for the enhanced energy deposition enhancement found in the previous subsection?
To reveal the origin of the energy deposition enhancement under double-color laser fields, we address to $\eta$, $\hbar\omega_1$, and $\hbar\omega_2$ dependence of these two quantities.
The number of excited electrons $n_{\mathrm{ex}}$ is evaluated as,
\begin{equation}
    n_{\mathrm{ex}}=\sum_{i\in \mathrm{cond}}2\left|\int\dd \mathbf{r}\sum_j\phi_i^*(\mathbf{r},0)\phi_j(\mathbf{r},T+t_{\mathrm{delay}})\right|^2,
\end{equation}
and the absorbed energy per carrier $E_{\mathrm{mean}}$ as,
\begin{equation}
    E_{\mathrm{mean}}=\frac{W}{n_{\mathrm{ex}}}.
\end{equation}

\begin{figure}[tb]
  \centering
  \includegraphics[keepaspectratio,width=\hsize]{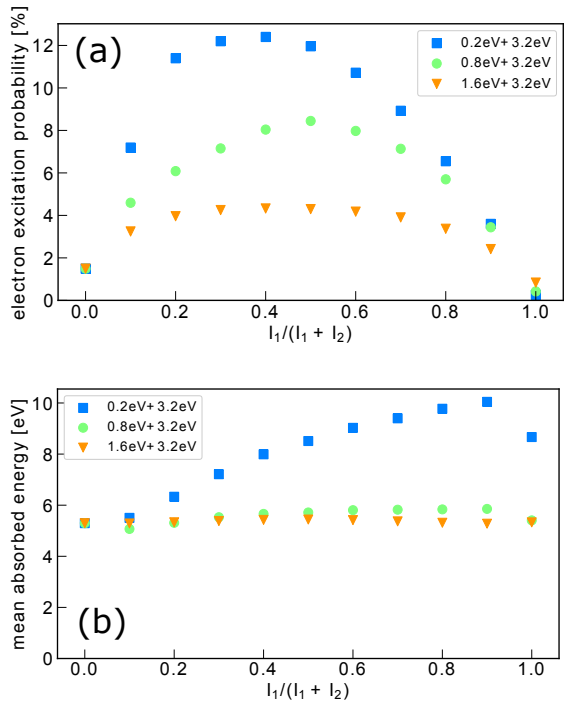}
  \caption[avenex-omega1]{(a) Fraction of excited electrons and (b) mean absorbed energy per excited electron as a function of mixing ratio $\eta$ under $I_{\rm tot}=10^{12}$ W/$\mathrm{cm}^2$ for three different combinations of $\hbar\omega_1$ and $\hbar\omega_2=3.2$ eV indicated in the legend. 
  }
  \label{avenex-omega1}
\end{figure}

Figure~\ref{avenex-omega1} shows $n_{\mathrm{ex}}$ and $E_{\mathrm{mean}}$ as functions of the mixing ratio $\eta$ for $\hbar\omega_2=3.2\,{\rm eV}$ and three different values of $\hbar\omega_1$.
We can see that $n_{\mathrm{ex}}$ follows the trend in Fig.~\ref{omega1}, strongly enhanced by dual-color irradiation, compared with the single-color cases ($\eta=0,1$), and peaking at $\eta\sim 0.5$ [Fig.~\ref{avenex-omega1}(a)].
The enhancement of interband excitation is larger for smaller $\hbar \omega_1$, again consistent with the trend in Fig.~\ref{omega1}.
On the other hand, the absorbed energy per carrier $E_{\mathrm{mean}}$ is surprisingly nearly independent of $\eta$ for $\hbar\omega_1=0.8$ and 1.6 eV [Fig.~\ref{avenex-omega1}(b)]. 
\red{Interestingly, $E_{\mathrm{mean}}$ is larger than the direct bandgap (3.1 eV), because not only the lowest but also higher conduction bands are populated (see Appendix \ref{sec:appendixA}).}
For the case of $\hbar\omega_1=0.2$ eV, although $E_{\mathrm{mean}}$ gradually increases with $\eta$,
it reaches maximum at $\eta\sim 0.9$, rather than $\sim 0.5$.
Furthermore, the enhancement factor (two at most) is much smaller than that of $W$ and $n_{\mathrm{ex}}$.

\begin{figure}[tb]
  \centering
  \includegraphics[keepaspectratio,width=\hsize]{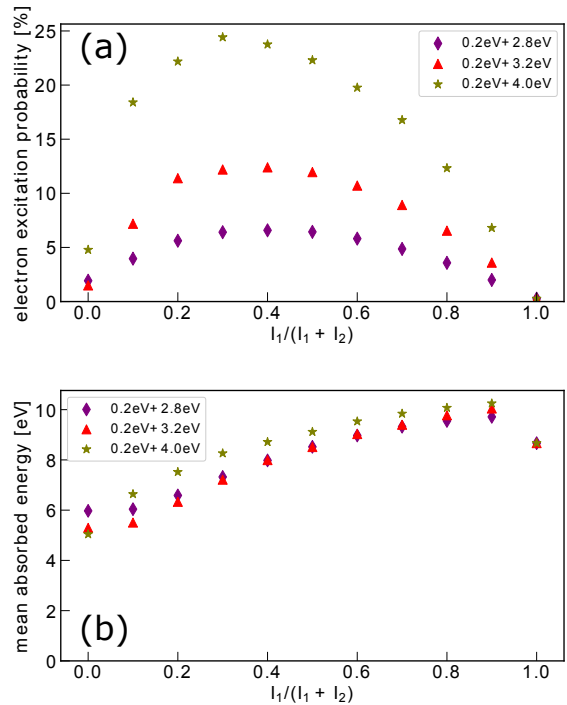}
  \caption[avenex-omega2]{(a) Fraction of excited electrons and (b) mean absorbed energy per excited electron as a function of mixing ratio $\eta$ under $I_{\rm tot}=10^{12}$ W/$\mathrm{cm}^2$ for three different combinations of $\hbar\omega_1=0.2$ eV and $\hbar\omega_2$ indicated in the legend.}
  \label{avenex-omega2}
\end{figure}

Figure~\ref{avenex-omega2} displays $n_{\mathrm{ex}}$ [panel (a)] and $E_{\mathrm{mean}}$ vs.~$\eta$ [panel (b)]
shows $n_{\mathrm{ex}}$ for $\hbar\omega_1=0.2\,{\rm eV}$ and three different values of $\hbar\omega_2$.
Again, the number of excited electrons $n_{\mathrm{ex}}$ follows a trend similar to that in Fig.~\ref{omega2}, peaking at comparable mixing of the two colors and increasing with $\hbar\omega_2$.
In contrast, the mean absorbed energy $E_{\mathrm{mean}}$ is nearly independent of $\hbar\omega_2$ and only a weak function of $\eta$.

Hence, the energy absorption enhanced by two-color pulse irradiation originates mainly from increase in excitation from the valence to the conduction band, rather than further excitation of carriers generated in the conduction band.
These findings suggest the following mechanism underlying the increased energy absorption under two-color laser fields.
The shorter-wavelength component $\hbar\omega_2$, if alone, would excite only valence electrons of crystal momenta $\mathbf{k}_{\rm res}$ where $\hbar\omega_2$ is resonant with the excitation energy.
In two-color cases, however, the longer-wavelength component $\hbar\omega_1$ drives the intraband motion in the {\it valence} band; the electronic momentum is shifted as $\mathbf{k}(t)=\mathbf{k}_0+\mathbf{A}_1(t)$, with $\mathbf{k}_0$ being its field-free value.
The peak amplitude of the vector potential $\mathbf{A}_1(t)$ is 0.06431, 0.1286, and 0.5145 atomic units (a.u.) for $\hbar\omega_1=1.6$, 0.8, 0.2 eV, respectively, at $I_1=5\times 10^{11} \mathrm{W/cm^2}$.
These values are to be compared with the half width of the first Brillouin zone along $\Gamma X, \  0.304$ a.u.
Thus, a substantial part of valence electrons can pass by the resonant momenta $\mathbf{k}_{\rm res}$ during the intraband motion, enabling much more electrons get excited. From the viewpoint of $\hbar\omega_1$, on the other hand, it is much smaller than the band gap, so that excitation is not much induced by the longer-wavelength component alone but facilitated by mixing the shorter-wavelength component $\hbar\omega_2$.
The interplay and balance between the intraband motion by $\hbar\omega_1$ (through vector potential) and interband excitation by $\hbar\omega_2$ (through photon energy) leads to increase in generated carriers, as in Figs.~\ref{avenex-omega1}(a) and \ref{avenex-omega2}(a),
\red{and to maximum enhancement of energy absorption at approximately equal two-color mixing.}
Interestingly, for the case of $\hbar\omega_1=0.2$ eV, the maximum momentum shift (0.5145 a.u.) given by the vector potential amplitude exceeds the Brillouin zone radius. Hence, excited electrons can be shifted in the conduction to reach the Brillouin zone edge and then further excited to upper bands, which results in increased mean absorbed energy [Figs.~\ref{avenex-omega1}(b) and \ref{avenex-omega2}(b)].
It should be emphasized, nevertheless, that the intraband motion in the {\it valence} band plays a dominant role in the enhanced energy absorption, rather than that in the {\it conduction} band.

\subsection{\label{subsec:population}\red{P}opulation analysis in $k$-space}
To further verify the above-proposed mechanism, we analyze the time-dependent $k$-resolved population defined by,
\begin{equation}
    \label{eq:population}
    \rho_{\mathbf{k},i}(t)=\left|\int\dd \mathbf{r}\sum_{j\in \mathrm{occ}}\phi_{\mathbf{k}+e\mathbf{A}(t),i}^*(\mathbf{r},0)\phi_{\mathbf{k},j}(\mathbf{r},t)\right|^2,
\end{equation}
where $i$ and $j$ denote the band index. We consider photon energy combination $\hbar\omega_1=1.6$ eV and $\hbar\omega_2=3.2$ eV ($I_{\rm tot}=10^{12}$ W/$\mathrm{cm}^2$) and focus on the following three $k$-points:
$\Gamma \ (\mathbf{k}=[-0.019,-0.019,-0.019])$, $\Gamma_1 \ (\mathbf{k}=[0.019, -0.019, 0.057])$, and P $(\mathbf{k}=[-0.019, -0.019, -0.095])$. 
\begin{figure}[tb]
  \centering
  \includegraphics[keepaspectratio,width=\hsize]{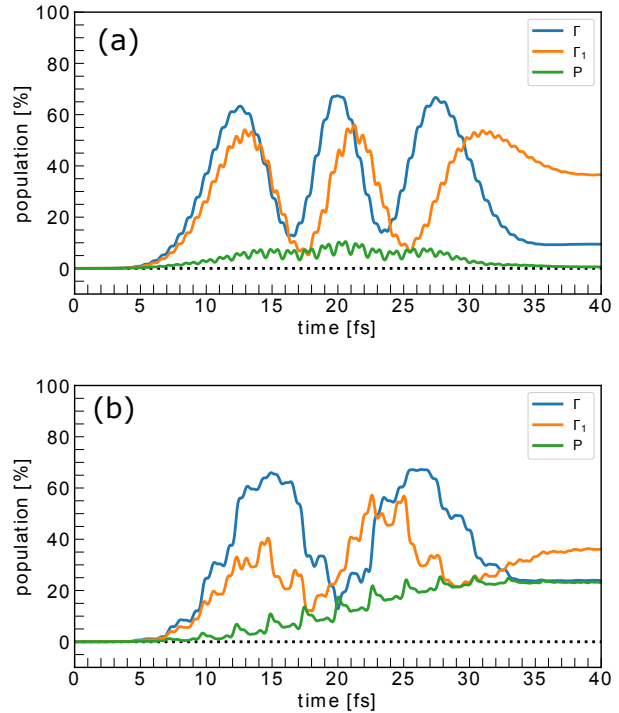}
  \caption[KpopT40]{Temporal evolution of the conduction band population at the $\Gamma$, $\Gamma_1$, and P points for (a) $\eta=0$ and (b) $\eta=0.5$ under $I_{\rm tot}=10^{12}$ W/$\mathrm{cm}^2$.
  }
  \label{KpopT40}
\end{figure}

\begin{figure}[tb]
  \centering
  \includegraphics[keepaspectratio,width=\hsize]{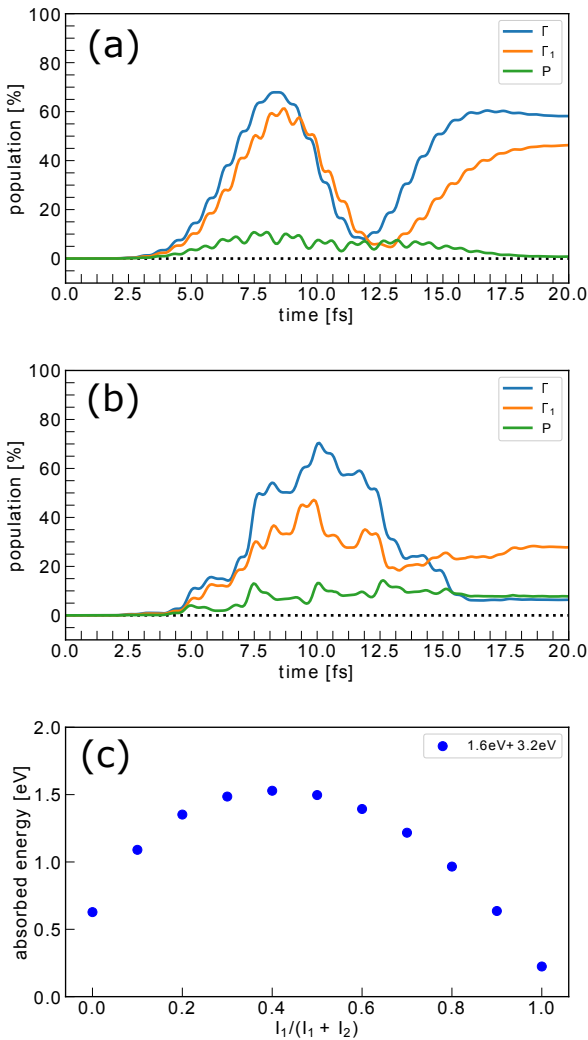}
  \caption[T20]{Temporal evolution of the conduction band population at the $\Gamma$, $\Gamma_1$, and P points for (a) $\eta=0$ and (b) $\eta=0.5$ for the case of 7.2 fs FWHM pulse width. 
  (c) absorbed energy vs. IR mixing ratio $\eta$ for the case of 7.2 fs FWHM pulse width. The total peak intensity is $I_{\rm tot}=10^{12}$ W/$\mathrm{cm}^2$.}
  \label{T20}
\end{figure}

\begin{figure}[tb]
  \centering
  \includegraphics[keepaspectratio,width=\hsize]{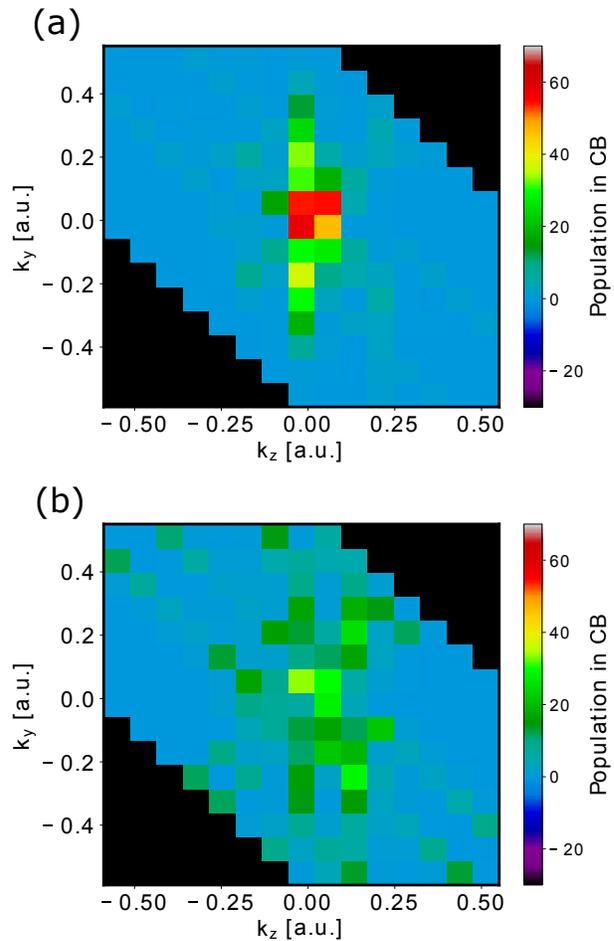}
  \caption[T20map]{The residual conduction band population on a cross section ($k_x=-0.019$ a.u.) for (a) $\eta=0$ and (b) $\eta=0.5$ under the short pulse (7.2 fs in FWHM, $I_{\rm tot}=10^{12}$ W/$\mathrm{cm}^2$). Negative value is assigned outside the first Brillouin zone.}
  \label{T20map}
\end{figure}

For the case of $\eta=0.0$ [Fig.~\ref{KpopT40}(a)], the population at and near the $\Gamma$ point ($\Gamma,\Gamma_1$) exhibits Rabi flopping \red{\cite{Zhao1996-ua,Wismer2016-hi}}, which means the excitation around the $\Gamma$ point is saturated, while almost no electrons are excited at the P point, which is far from the $\Gamma$ point. 
Thus, the energy transfer takes place only around the $\Gamma$ point, and even there, the laser pulse energy is ``wasted". 
For $\eta=0.5$ [Fig.~\ref{KpopT40}(b)], on the other hand, while the conduction band is still populated at $\Gamma$ and $\Gamma_1$ as much as for $\eta=0$, the P point electron is comparably excited. 
This observation indeed supports our discussion in the previous subsection that mixing the lower photon energy component into the driving laser field induces intraband motion in the valence band, expressed by subscript $\mathbf{k}+e\mathbf{A}(t)$ in Eq.~\eqref{eq:population}, and broadens the excitable crystal momentum range, leading to more efficient use of the laser energy.
Whereas we also see increase in residual population even at the $\Gamma$ point, it happens accidentally during the Rabi oscillation.
Indeed, under a shorter pulse (7.2 fs in FWHM, $I_{\rm tot}=10^{12}$ W/$\mathrm{cm}^2$), the final population of the $\Gamma$ and $\Gamma_1$ points for $\eta=0.5$ [Fig.~\ref{T20}(b)] is smaller than for $\eta=0$ [Fig.~\ref{T20}(a)].
Nevertheless, the absorbed energy is still maximized at $\eta\sim0.5$ [Fig.~\ref{T20}(c)].
\red{In Fig.~\ref{T20map}, we compare the residual population distributions in the conduction band shown as a sectional
view along $k_x=-0.019$ a.u. for $\eta=0.0$ and $\eta=0.5$ after the short laser pulse (7.2 fs in FWHM, $I_{\rm tot}=10^{12}$ W/$\mathrm{cm}^2$) irradiation. Again, we find that wider area of the $k$ space is excited
for $\eta=0.5$ [Fig.~\ref{T20map}(b)] than for $\eta=0.0$ [Fig.~\ref{T20map}(a)], leading to higher total excitation probability.}
These observations indicate that not the population transfer at the $\Gamma$ point alone but the excitable $\mathbf{k}$ range expanded by the addition of the longer-wavelength component has a major contribution to the enhanced energy absorption.

\section{\label{sec:conclusions} Conclusions}
We have investigated energy absorption by crystalline silicon under two-color laser pulse irradiation, by varying the mixing ratio with the total intensity or, equivalently, fluence fixed constant.
We have specifically considered the higher photon energy $\hbar\omega_2$ (3.2 eV) in the UV range above the optical gap energy $E_g$ and the lower one $\hbar\omega_1$ (0.2-1.6 eV) in the infrared below it. 
Our TDDFT computation have shown that energy transfer is substantially enhanced by simultaneous two-color irradiation and maximized when both color components are roughly equally mixed.
The longer the wavelength of the longer-wavelength component, the more significant the effect.
Increase in carrier generation, rather than that in absorbed energy per carrier, accounts for increase in the total absorbed energy.
All these observations, along with our $k$-resolved population analysis, have revealed that the electron dynamics in the valence band driven by the longer-wavelength component $\hbar\omega_1$ plays a crucial role, in contrast to previous studies, which have emphasized the interplay of different multiphoton paths or the dynamics of the generated carriers.
The $\hbar\omega_2$ field induces interband excitation (carrier injection) around the $\Gamma$ point, where the transition energy is resonant with $\hbar\omega_2$. On the other hand, the $\hbar\omega_1$ field of a large vector potental amplitude (corresponding to the maximum momentum shift) drives intraband motion in the valence band, enabling electrons initially far from the $\Gamma$ point to pass through the resonance, thus, extending the excitable $k$ range. Additionally, for the case of $\hbar\omega_1=0.2$ eV, the induced momentum shift exceeds the width of Brillouin zone along $\Gamma X$ and promotes further excitation of carrier electrons to upper conduction bands.

\red{The decoherence would have some effects depending on dephasing time $T_2$. 
\citet{Floss2018-yj} have pointed out that $T_2$ is order of 10 femtoseconds, which is longer than the time scale of Rabi oscillation observed in our calculation. \citet{Freeman2022-dn} have recently shown that relaxation effect on high-harmonic generation emerges only after several tens of femtoseconds. Hence, our simulation results are semi-quantitatively valid even in the presence of the decoherence effect.}

\redd{The degree of ionization and energy per electron are rather high for the photon energy combination (0.2 eV, 3.2 eV). Thus, we should be cautious if we discussed phenomena at longer time scale, which eventually leads to material processing. On the other hand, in this work, we limit ourselves to the ultrashort time scale before the lattice starts to move significantly.
Moreover, the photon energy combination (1.6 eV, 3.2 eV), for which the degree of ionization is 4\%, is used in the main part of our analysis. The Rabi oscillation observed in Fig.~\ref{KpopT40} implies that the bandgap is close to the original one, thus, indicates that our discussion of the ultrafast electron excitation dynamics based on the ground state band structure is approximately valid and useful for physical understanding of the early stage of laser material processing.}

The findings of this work, suggesting dramatic improvement of femtosecond laser ablation rate of bandgap materials, could be experimentally examined, using the combination of ultrashort intense mid-infrared (MIR) and ultraviolet laser pulses. 
Thereby, increase in generated carrier will be easier to probe directly rather than absorbed energy itself.
Use of a terahertz pulse instead of IR may be even more advantageous, since its vector potential has an even larger amplitude.

\section*{\label{sec:acknowledgements} ACKNOWLEDGEMENTS}
This research was supported by MEXT Quantum Leap Flagship Program (MEXT Q-LEAP) Grant Number JPMXS0118067246. 
This research was also partially supported by JSPS KAKENHI Grant Number 20H05670, 18K14145, 19H02623, JST COI Grant Number JPMJCE1313, and JST CREST under Grant Number JPMJCR16N5, and the Exploratory Challenge on Post-K Computer from MEXT.. 
M.T. gratefully acknowledges support from the Graduate School of Engineering, The University of Tokyo, Graduate Student Special Incentives Program. M.T. also gratefully thanks support through crowd funding platform \it academist \rm by Ryosuke Shibato, Hitomi Suto, Shunsuke A. Sato, Yusaku Karibe, Shion Chen. The numerical calculations are partially performed on supercomputer Oakbridge-CX (the University of Tokyo) and the K computer provided by the RIKEN Advanced 
Institute for Computational Science and Oakforest-PACS by JCAHPC through the HPCI System Research 
project (Project ID: 190147).
\appendix
\def\thesection{\Alph{section}}

\section{\label{sec:appendixC} Crystal structure and band structure}
\begin{figure}[tb]
  \centering
  \includegraphics[keepaspectratio,width=\hsize]{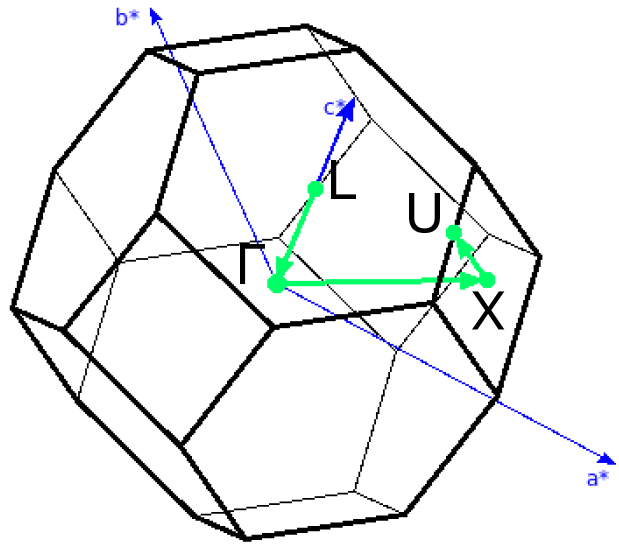}
  \caption[cell]{Wigner-Seitz cell of the crystalline silicon in the reciprocal space. The image is generated using XCrySDen \cite{Kokalj2003-nf}.}
  \label{cell}
\end{figure}

\begin{figure}[tb]
  \centering
  \includegraphics[keepaspectratio,width=\hsize]{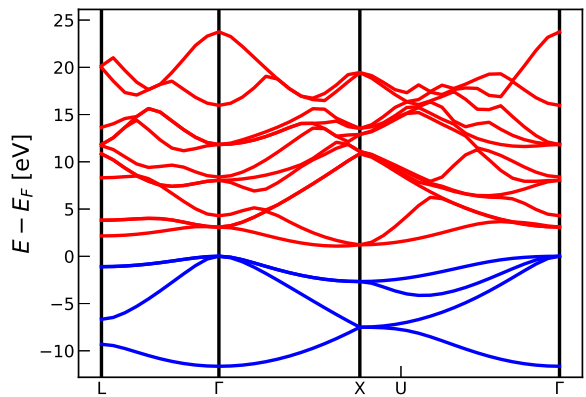}
  \caption[bands]{Energy dispersion for the conduction and valence bands of silicon, calculated using ABINIT code \cite{Gonze2020-yq,Romero2020-hq} with the lattice constant $a=5.468 $ \AA. Red and blue lines indicate condution and valence bands respectively.}
  \label{bands}
\end{figure}

Figure~\ref{cell} shows the Wigner-Seitz cell of the crystalline silicon in reciprocal space. Figure~\ref{bands} shows the energy dispersion for the silicon calculated using ABINIT code \cite{Gonze2020-yq,Romero2020-hq} with the lattice constant $a=5.468 $ \AA.

\section{\label{sec:appendixB}Convergence with respect to grid spacings and time step}
\begin{figure}[tb]
  \centering
  \includegraphics[keepaspectratio,width=\hsize]{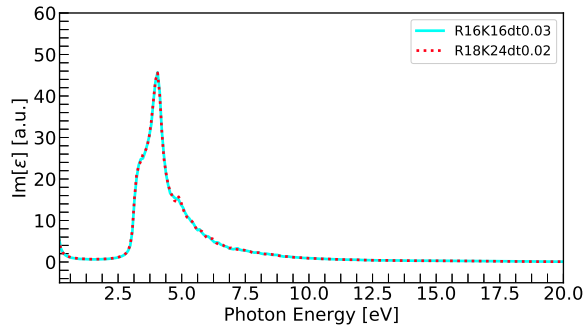}
  \caption[conv]{Imaginary part of the dielectric function using our parameters (light blue solid line) and finer parameters (red dashed line) for real-space, $k$-space, and time discretization. See text for the parameter values.}
  \label{conv}
\end{figure}

Figure~\ref{conv} compares the imaginary part of the dielectric function $\Im[\epsilon]$ calculated with two different real($k$)-space grid widths and time steps. Red dashed line indicates $\Im[\epsilon]$ with $16^3$ grid points in real-space and $k$-space, blue line with $18^3$ grid points in real-space and $24^3$ grid points in k-space. The time step is 0.03 a.u. with the former parameter set, 0.02 a.u. with the latter. Both results are perfectly overlapped, indicating the simulation results are well-converged.

\section{\label{sec:appendixA} Excited carrier population in each conduction band}
\begin{figure}[tb]
  \centering
  \includegraphics[keepaspectratio,width=\hsize]{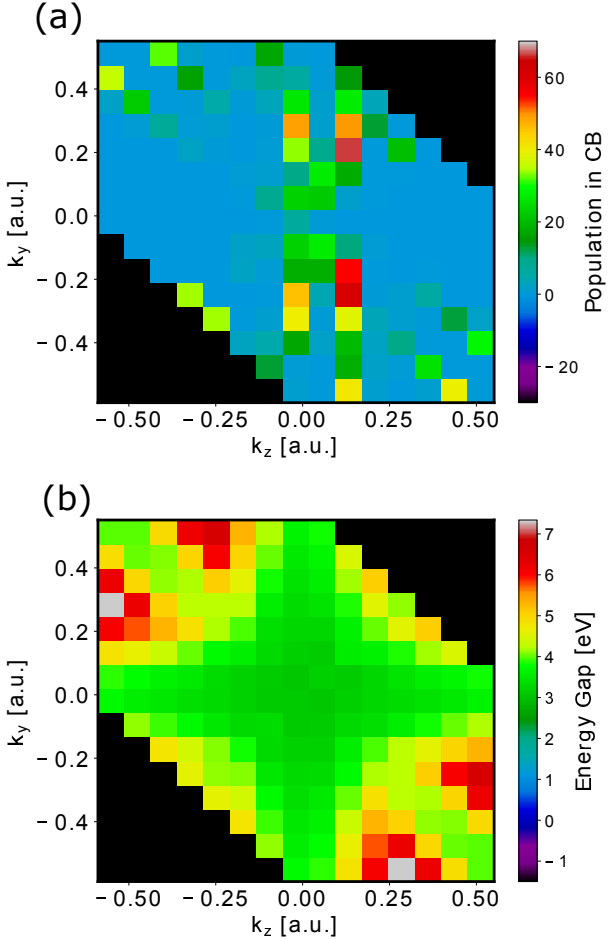}
  \caption[CB4]{(a) The residual population of the lowest conduction band on a cross section ($k_x=-0.019$ a.u.) after irradiated by the short pulse (7.2 fs in FWHM, $I_{\rm tot}=10^{12}$ W/$\mathrm{cm}^2$, $\eta=0.5$). (b) Energy gap between the lowest conduction band and the highest valence band. Negative value is assigned outside the first Brillouin zone.}
  \label{CB4}
\end{figure}

\begin{figure}[tb]
  \centering
  \includegraphics[keepaspectratio,width=\hsize]{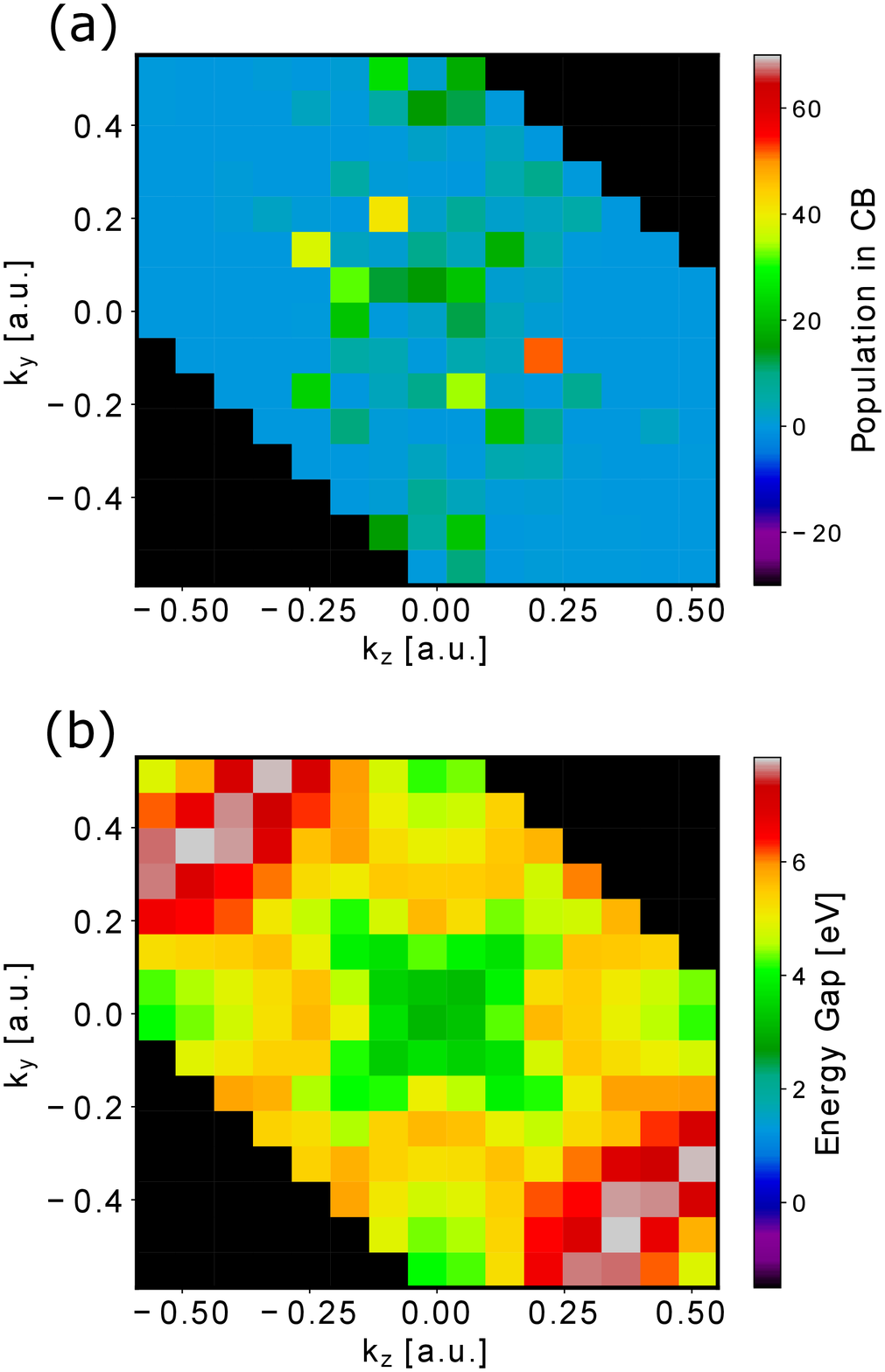}
  \caption[CB5]{(a) The residual population of the 2nd lowest conduction band on a cross section ($k_x=-0.019$ a.u.) after irradiated by the short pulse (7.2 fs in FWHM, $I_{\rm tot}=10^{12}$ W/$\mathrm{cm}^2$, $\eta=0.5$). (b) Energy gap between the 2nd lowest conduction band and the highest valence band. Negative value is assigned outside the first Brillouin zone.}
  \label{CB5}
\end{figure}

\begin{figure}[tb]
  \centering
  \includegraphics[keepaspectratio,width=\hsize]{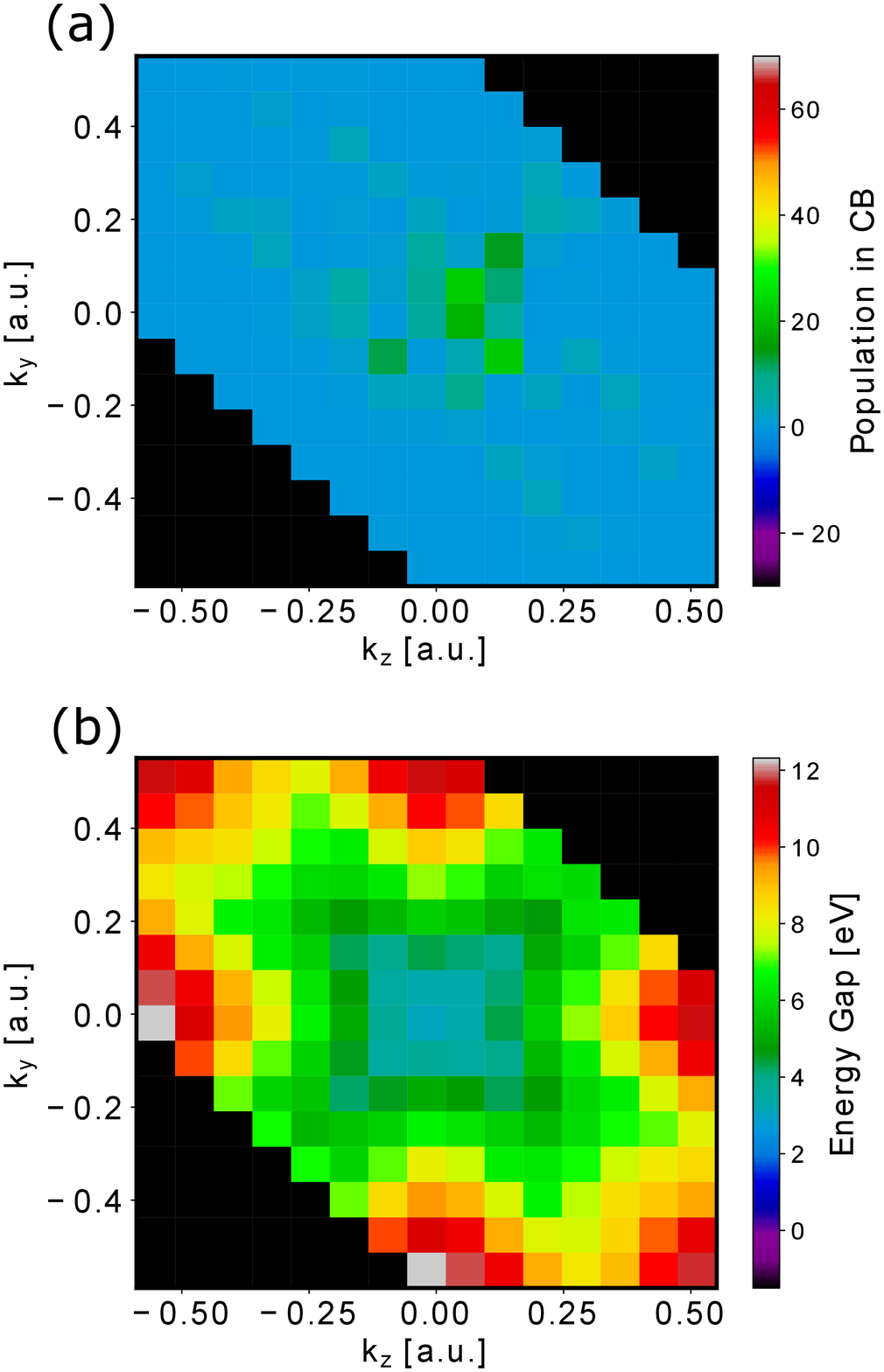}
  \caption[CB6]{(a) The residual population of the 3rd lowest conduction band on a cross section ($k_x=-0.019$ a.u.) after irradiated by the short pulse (7.2 fs in FWHM, $I_{\rm tot}=10^{12}$ W/$\mathrm{cm}^2$, $\eta=0.5$). (b) Energy gap between the 3rd lowest conduction band and the highest valence band. Negative value is assigned outside the first Brillouin zone.}
  \label{CB6}
\end{figure}

\begin{figure}[tb]
  \centering
  \includegraphics[keepaspectratio,width=\hsize]{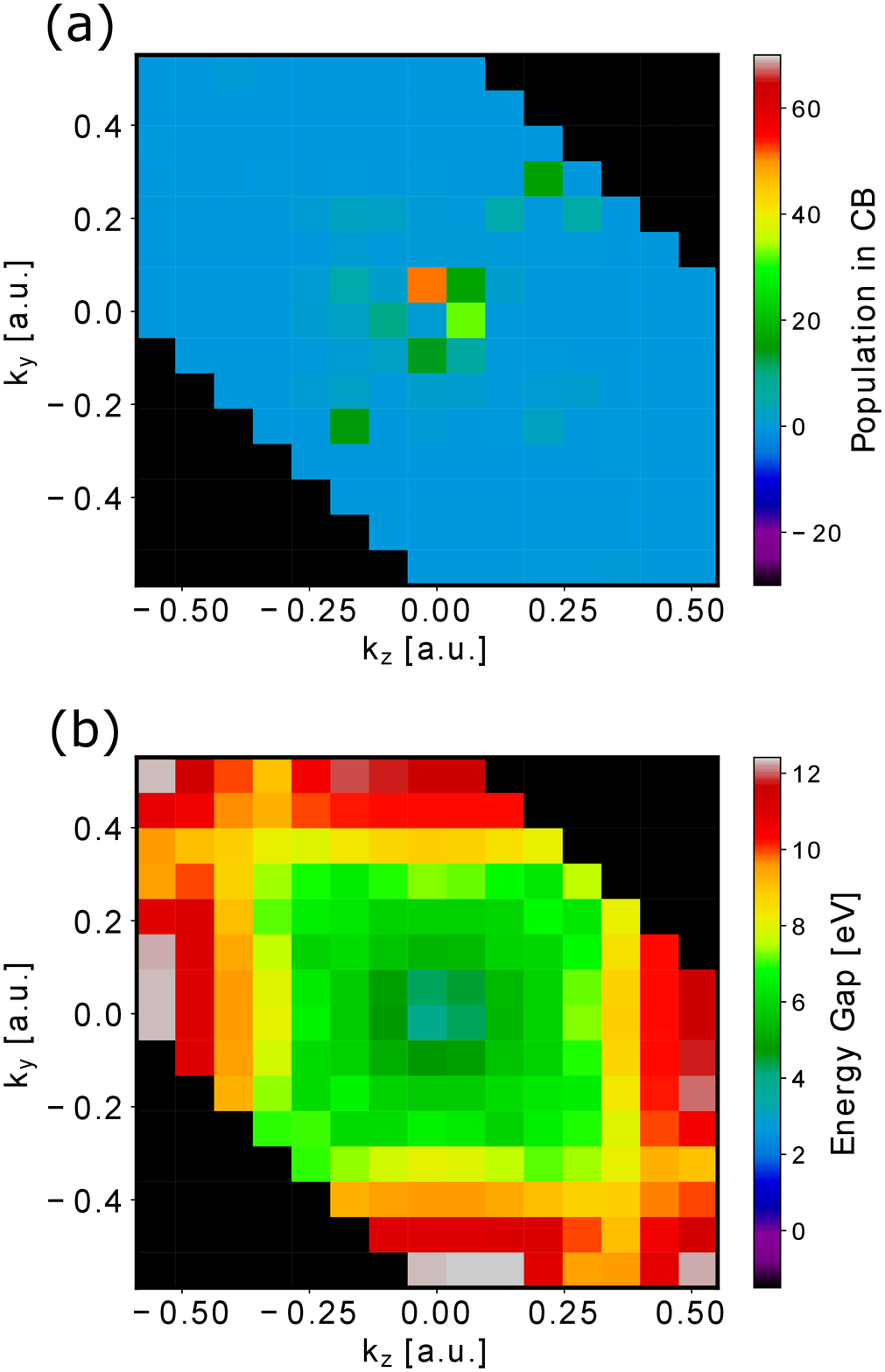}
  \caption[CB7]{(a) The residual population of the 4th lowest conduction band on a cross section ($k_x=-0.019$ a.u.) after irradiated by the short pulse (7.2 fs in FWHM, $I_{\rm tot}=10^{12}$ W/$\mathrm{cm}^2$, $\eta=0.5$). (b) Energy gap between the 4th lowest conduction band and the highest valence band. Negative value is assigned outside the first Brillouin zone.}
  \label{CB7}
\end{figure}

Figures~\ref{CB4}-\ref{CB7} show the residual population map of each conduction band within the same cross section as Fig.~\ref{T20map} for $\eta=0.5$ under $\hbar\omega_1=1.6$ eV and $\hbar\omega_2=3.2$ eV ($I_1+I_2=10^{12} \mathrm{W}/\mathrm{cm}^2$) and the energy gap between the valence top band and each conduction band. We see that the electrons are excited around the minimum bandgap (3.1 eV) but also at $k$ points where the energy gap is larger than it. Therefore, the mean absorbed energy exceeds the direct bandgap (3.1 eV) [see Fig.~\ref{avenex-omega1}(b)].


\begin{thebibliography}{72}%
\makeatletter
\providecommand \@ifxundefined [1]{%
 \@ifx{#1\undefined}
}%
\providecommand \@ifnum [1]{%
 \ifnum #1\expandafter \@firstoftwo
 \else \expandafter \@secondoftwo
 \fi
}%
\providecommand \@ifx [1]{%
 \ifx #1\expandafter \@firstoftwo
 \else \expandafter \@secondoftwo
 \fi
}%
\providecommand \natexlab [1]{#1}%
\providecommand \enquote  [1]{``#1''}%
\providecommand \bibnamefont  [1]{#1}%
\providecommand \bibfnamefont [1]{#1}%
\providecommand \citenamefont [1]{#1}%
\providecommand \href@noop [0]{\@secondoftwo}%
\providecommand \href [0]{\begingroup \@sanitize@url \@href}%
\providecommand \@href[1]{\@@startlink{#1}\@@href}%
\providecommand \@@href[1]{\endgroup#1\@@endlink}%
\providecommand \@sanitize@url [0]{\catcode `\\12\catcode `\$12\catcode
  `\&12\catcode `\#12\catcode `\^12\catcode `\_12\catcode `\%12\relax}%
\providecommand \@@startlink[1]{}%
\providecommand \@@endlink[0]{}%
\providecommand \url  [0]{\begingroup\@sanitize@url \@url }%
\providecommand \@url [1]{\endgroup\@href {#1}{\urlprefix }}%
\providecommand \urlprefix  [0]{URL }%
\providecommand \Eprint [0]{\href }%
\providecommand \doibase [0]{https://doi.org/}%
\providecommand \selectlanguage [0]{\@gobble}%
\providecommand \bibinfo  [0]{\@secondoftwo}%
\providecommand \bibfield  [0]{\@secondoftwo}%
\providecommand \translation [1]{[#1]}%
\providecommand \BibitemOpen [0]{}%
\providecommand \bibitemStop [0]{}%
\providecommand \bibitemNoStop [0]{.\EOS\space}%
\providecommand \EOS [0]{\spacefactor3000\relax}%
\providecommand \BibitemShut  [1]{\csname bibitem#1\endcsname}%
\let\auto@bib@innerbib\@empty
\bibitem [{\citenamefont {Chichkov}\ \emph {et~al.}(1996)\citenamefont
  {Chichkov}, \citenamefont {Momma}, \citenamefont {Nolte}, \citenamefont
  {Von~Alvensleben},\ and\ \citenamefont {T{\"u}nnermann}}]{Tiinnermann}%
  \BibitemOpen
  \bibfield  {author} {\bibinfo {author} {\bibfnamefont {B.~N.}\ \bibnamefont
  {Chichkov}}, \bibinfo {author} {\bibfnamefont {C.}~\bibnamefont {Momma}},
  \bibinfo {author} {\bibfnamefont {S.}~\bibnamefont {Nolte}}, \bibinfo
  {author} {\bibfnamefont {F.}~\bibnamefont {Von~Alvensleben}},\ and\ \bibinfo
  {author} {\bibfnamefont {A.}~\bibnamefont {T{\"u}nnermann}},\ }\bibfield
  {title} {\bibinfo {title} {Femtosecond, picosecond and nanosecond laser
  ablation of solids},\ }\href@noop {} {\bibfield  {journal} {\bibinfo
  {journal} {Applied physics A}\ }\textbf {\bibinfo {volume} {63}},\ \bibinfo
  {pages} {109} (\bibinfo {year} {1996})}\BibitemShut {NoStop}%
\bibitem [{\citenamefont {Kerse}\ \emph {et~al.}(2016)\citenamefont {Kerse},
  \citenamefont {Kalayc{\i}o{\u{g}}lu}, \citenamefont {Elahi}, \citenamefont
  {{\c{C}}etin}, \citenamefont {Kesim}, \citenamefont {Ak{\c{c}}aalan},
  \citenamefont {Yava{\c{s}}}, \citenamefont {A{\c{s}}{\i}k}, \citenamefont
  {{\"O}ktem}, \citenamefont {Hoogland}, \citenamefont {Holzwarth},\ and\
  \citenamefont {Ilday}}]{Ilday}%
  \BibitemOpen
  \bibfield  {author} {\bibinfo {author} {\bibfnamefont {C.}~\bibnamefont
  {Kerse}}, \bibinfo {author} {\bibfnamefont {H.}~\bibnamefont
  {Kalayc{\i}o{\u{g}}lu}}, \bibinfo {author} {\bibfnamefont {P.}~\bibnamefont
  {Elahi}}, \bibinfo {author} {\bibfnamefont {B.}~\bibnamefont {{\c{C}}etin}},
  \bibinfo {author} {\bibfnamefont {D.~K.}\ \bibnamefont {Kesim}}, \bibinfo
  {author} {\bibfnamefont {{\"O}.}~\bibnamefont {Ak{\c{c}}aalan}}, \bibinfo
  {author} {\bibfnamefont {S.}~\bibnamefont {Yava{\c{s}}}}, \bibinfo {author}
  {\bibfnamefont {M.~D.}\ \bibnamefont {A{\c{s}}{\i}k}}, \bibinfo {author}
  {\bibfnamefont {B.}~\bibnamefont {{\"O}ktem}}, \bibinfo {author}
  {\bibfnamefont {H.}~\bibnamefont {Hoogland}}, \bibinfo {author}
  {\bibfnamefont {R.}~\bibnamefont {Holzwarth}},\ and\ \bibinfo {author}
  {\bibfnamefont {F.~{\"O}.}\ \bibnamefont {Ilday}},\ }\bibfield  {title}
  {\bibinfo {title} {Ablation-cooled material removal with ultrafast bursts of
  pulses},\ }\href@noop {} {\bibfield  {journal} {\bibinfo  {journal} {Nature}\
  }\textbf {\bibinfo {volume} {537}},\ \bibinfo {pages} {84} (\bibinfo {year}
  {2016})}\BibitemShut {NoStop}%
\bibitem [{\citenamefont {Liu}\ \emph {et~al.}(1997)\citenamefont {Liu},
  \citenamefont {Du},\ and\ \citenamefont {Mourou}}]{Mourou}%
  \BibitemOpen
  \bibfield  {author} {\bibinfo {author} {\bibfnamefont {X.}~\bibnamefont
  {Liu}}, \bibinfo {author} {\bibfnamefont {D.}~\bibnamefont {Du}},\ and\
  \bibinfo {author} {\bibfnamefont {G.}~\bibnamefont {Mourou}},\ }\bibfield
  {title} {\bibinfo {title} {Laser ablation and micromachining with ultrashort
  laser pulses},\ }\href@noop {} {\bibfield  {journal} {\bibinfo  {journal}
  {IEEE journal of quantum electronics}\ }\textbf {\bibinfo {volume} {33}},\
  \bibinfo {pages} {1706} (\bibinfo {year} {1997})}\BibitemShut {NoStop}%
\bibitem [{\citenamefont {Gattass}\ and\ \citenamefont {Mazur}(2008)}]{Mazur}%
  \BibitemOpen
  \bibfield  {author} {\bibinfo {author} {\bibfnamefont {R.~R.}\ \bibnamefont
  {Gattass}}\ and\ \bibinfo {author} {\bibfnamefont {E.}~\bibnamefont
  {Mazur}},\ }\bibfield  {title} {\bibinfo {title} {Femtosecond laser
  micromachining in transparent materials},\ }\href@noop {} {\bibfield
  {journal} {\bibinfo  {journal} {Nature photonics}\ }\textbf {\bibinfo
  {volume} {2}},\ \bibinfo {pages} {219} (\bibinfo {year} {2008})}\BibitemShut
  {NoStop}%
\bibitem [{\citenamefont {Ostrikov}\ \emph {et~al.}(2016)\citenamefont
  {Ostrikov}, \citenamefont {Beg},\ and\ \citenamefont {Ng}}]{Ostrikov2016-sf}%
  \BibitemOpen
  \bibfield  {author} {\bibinfo {author} {\bibfnamefont {K.}~\bibnamefont
  {Ostrikov}}, \bibinfo {author} {\bibfnamefont {F.}~\bibnamefont {Beg}},\ and\
  \bibinfo {author} {\bibfnamefont {A.}~\bibnamefont {Ng}},\ }\bibfield
  {title} {\bibinfo {title} {Colloquium: Nanoplasmas generated by intense
  radiation},\ }\href@noop {} {\bibfield  {journal} {\bibinfo  {journal}
  {Review of Modern Physics}\ }\textbf {\bibinfo {volume} {88}},\ \bibinfo
  {pages} {011001} (\bibinfo {year} {2016})}\BibitemShut {NoStop}%
\bibitem [{\citenamefont {Sugioka}\ and\ \citenamefont
  {Cheng}(2014)}]{Sugioka2014-rb}%
  \BibitemOpen
  \bibfield  {author} {\bibinfo {author} {\bibfnamefont {K.}~\bibnamefont
  {Sugioka}}\ and\ \bibinfo {author} {\bibfnamefont {Y.}~\bibnamefont
  {Cheng}},\ }\bibfield  {title} {\bibinfo {title} {Ultrafast lasers---reliable
  tools for advanced materials processing},\ }\href@noop {} {\bibfield
  {journal} {\bibinfo  {journal} {Light: Science \& Applications}\ }\textbf
  {\bibinfo {volume} {3}},\ \bibinfo {pages} {e149} (\bibinfo {year}
  {2014})}\BibitemShut {NoStop}%
\bibitem [{\citenamefont {Sugioka}(2021)}]{Sugioka2021-fh}%
  \BibitemOpen
  \bibfield  {author} {\bibinfo {author} {\bibfnamefont {K.}~\bibnamefont
  {Sugioka}},\ }\bibfield  {title} {\bibinfo {title} {Will {GHz} burst mode
  create a new path to femtosecond laser processing?},\ }\href@noop {}
  {\bibfield  {journal} {\bibinfo  {journal} {International Journal of Extreme
  Manufacturing}\ }\textbf {\bibinfo {volume} {3}},\ \bibinfo {pages} {043001}
  (\bibinfo {year} {2021})}\BibitemShut {NoStop}%
\bibitem [{\citenamefont {Thorstensen}\ and\ \citenamefont
  {Erik~Foss}(2012)}]{Thorstensen2012-ji}%
  \BibitemOpen
  \bibfield  {author} {\bibinfo {author} {\bibfnamefont {J.}~\bibnamefont
  {Thorstensen}}\ and\ \bibinfo {author} {\bibfnamefont {S.}~\bibnamefont
  {Erik~Foss}},\ }\bibfield  {title} {\bibinfo {title} {Temperature dependent
  ablation threshold in silicon using ultrashort laser pulses},\ }\href@noop {}
  {\bibfield  {journal} {\bibinfo  {journal} {Journal of Applied Physics}\
  }\textbf {\bibinfo {volume} {112}},\ \bibinfo {pages} {103514} (\bibinfo
  {year} {2012})}\BibitemShut {NoStop}%
\bibitem [{\citenamefont {Rousse}\ \emph {et~al.}(2001)\citenamefont {Rousse},
  \citenamefont {Rischel}, \citenamefont {Fourmaux}, \citenamefont {Uschmann},
  \citenamefont {Sebban}, \citenamefont {Grillon}, \citenamefont {Balcou},
  \citenamefont {F{\"o}rster}, \citenamefont {Geindre}, \citenamefont
  {Audebert}, \citenamefont {Gauthier},\ and\ \citenamefont
  {Hulin}}]{Rousse2001-ws}%
  \BibitemOpen
  \bibfield  {author} {\bibinfo {author} {\bibfnamefont {A.}~\bibnamefont
  {Rousse}}, \bibinfo {author} {\bibfnamefont {C.}~\bibnamefont {Rischel}},
  \bibinfo {author} {\bibfnamefont {S.}~\bibnamefont {Fourmaux}}, \bibinfo
  {author} {\bibfnamefont {I.}~\bibnamefont {Uschmann}}, \bibinfo {author}
  {\bibfnamefont {S.}~\bibnamefont {Sebban}}, \bibinfo {author} {\bibfnamefont
  {G.}~\bibnamefont {Grillon}}, \bibinfo {author} {\bibfnamefont
  {P.}~\bibnamefont {Balcou}}, \bibinfo {author} {\bibfnamefont
  {E.}~\bibnamefont {F{\"o}rster}}, \bibinfo {author} {\bibfnamefont {J.~P.}\
  \bibnamefont {Geindre}}, \bibinfo {author} {\bibfnamefont {P.}~\bibnamefont
  {Audebert}}, \bibinfo {author} {\bibfnamefont {J.~C.}\ \bibnamefont
  {Gauthier}},\ and\ \bibinfo {author} {\bibfnamefont {D.}~\bibnamefont
  {Hulin}},\ }\bibfield  {title} {\bibinfo {title} {Non-thermal melting in
  semiconductors measured at femtosecond resolution},\ }\href@noop {}
  {\bibfield  {journal} {\bibinfo  {journal} {Nature}\ }\textbf {\bibinfo
  {volume} {410}},\ \bibinfo {pages} {65} (\bibinfo {year} {2001})}\BibitemShut
  {NoStop}%
\bibitem [{\citenamefont {Sundaram}\ and\ \citenamefont
  {Mazur}(2002)}]{Sundaram2002-gc}%
  \BibitemOpen
  \bibfield  {author} {\bibinfo {author} {\bibfnamefont {S.~K.}\ \bibnamefont
  {Sundaram}}\ and\ \bibinfo {author} {\bibfnamefont {E.}~\bibnamefont
  {Mazur}},\ }\bibfield  {title} {\bibinfo {title} {Inducing and probing
  non-thermal transitions in semiconductors using femtosecond laser pulses},\
  }\href@noop {} {\bibfield  {journal} {\bibinfo  {journal} {Nature Material}\
  }\textbf {\bibinfo {volume} {1}},\ \bibinfo {pages} {217} (\bibinfo {year}
  {2002})}\BibitemShut {NoStop}%
\bibitem [{\citenamefont {Shcheblanov}\ and\ \citenamefont
  {Povarnitsyn}(2016)}]{Shcheblanov2016-eu}%
  \BibitemOpen
  \bibfield  {author} {\bibinfo {author} {\bibfnamefont {N.~S.}\ \bibnamefont
  {Shcheblanov}}\ and\ \bibinfo {author} {\bibfnamefont {M.~E.}\ \bibnamefont
  {Povarnitsyn}},\ }\bibfield  {title} {\bibinfo {title} {Bond-breaking
  mechanism of vitreous silica densification by {IR} femtosecond laser
  pulses},\ }\href@noop {} {\bibfield  {journal} {\bibinfo  {journal}
  {Europhysics Letters}\ }\textbf {\bibinfo {volume} {114}},\ \bibinfo {pages}
  {26004} (\bibinfo {year} {2016})}\BibitemShut {NoStop}%
\bibitem [{\citenamefont {Wang}\ \emph {et~al.}(2011)\citenamefont {Wang},
  \citenamefont {Jiang}, \citenamefont {Wang}, \citenamefont {Li},
  \citenamefont {Yuan},\ and\ \citenamefont {Tsai}}]{Wang2011-pm}%
  \BibitemOpen
  \bibfield  {author} {\bibinfo {author} {\bibfnamefont {C.}~\bibnamefont
  {Wang}}, \bibinfo {author} {\bibfnamefont {L.}~\bibnamefont {Jiang}},
  \bibinfo {author} {\bibfnamefont {F.}~\bibnamefont {Wang}}, \bibinfo {author}
  {\bibfnamefont {X.}~\bibnamefont {Li}}, \bibinfo {author} {\bibfnamefont
  {Y.~P.}\ \bibnamefont {Yuan}},\ and\ \bibinfo {author} {\bibfnamefont
  {H.~L.}\ \bibnamefont {Tsai}},\ }\bibfield  {title} {\bibinfo {title}
  {First-principles calculations of the electron dynamics during femtosecond
  laser pulse train material interactions},\ }\href@noop {} {\bibfield
  {journal} {\bibinfo  {journal} {Physics Letters A}\ }\textbf {\bibinfo
  {volume} {375}},\ \bibinfo {pages} {3200} (\bibinfo {year}
  {2011})}\BibitemShut {NoStop}%
\bibitem [{\citenamefont {Tani}\ \emph {et~al.}(2021)\citenamefont {Tani},
  \citenamefont {Otobe}, \citenamefont {Shinohara},\ and\ \citenamefont
  {Ishikawa}}]{Tani2021-Se}%
  \BibitemOpen
  \bibfield  {author} {\bibinfo {author} {\bibfnamefont {M.}~\bibnamefont
  {Tani}}, \bibinfo {author} {\bibfnamefont {T.}~\bibnamefont {Otobe}},
  \bibinfo {author} {\bibfnamefont {Y.}~\bibnamefont {Shinohara}},\ and\
  \bibinfo {author} {\bibfnamefont {K.~L.}\ \bibnamefont {Ishikawa}},\
  }\bibfield  {title} {\bibinfo {title} {Semiclassical description of electron
  dynamics in extended systems under intense laser fields},\ }\href@noop {}
  {\bibfield  {journal} {\bibinfo  {journal} {Physical Review B}\ }\textbf
  {\bibinfo {volume} {104}},\ \bibinfo {pages} {075157} (\bibinfo {year}
  {2021})}\BibitemShut {NoStop}%
\bibitem [{\citenamefont {Sugioka}\ \emph {et~al.}(1993)\citenamefont
  {Sugioka}, \citenamefont {Wada}, \citenamefont {Tsunemi}, \citenamefont
  {Sakai}, \citenamefont {Takai}, \citenamefont {Moriwaki}, \citenamefont
  {Nakamura}, \citenamefont {Tashiro},\ and\ \citenamefont
  {Toyoda}}]{Sugioka1993-hg}%
  \BibitemOpen
  \bibfield  {author} {\bibinfo {author} {\bibfnamefont {K.}~\bibnamefont
  {Sugioka}}, \bibinfo {author} {\bibfnamefont {S.}~\bibnamefont {Wada}},
  \bibinfo {author} {\bibfnamefont {A.}~\bibnamefont {Tsunemi}}, \bibinfo
  {author} {\bibfnamefont {T.}~\bibnamefont {Sakai}}, \bibinfo {author}
  {\bibfnamefont {H.}~\bibnamefont {Takai}}, \bibinfo {author} {\bibfnamefont
  {H.}~\bibnamefont {Moriwaki}}, \bibinfo {author} {\bibfnamefont
  {A.}~\bibnamefont {Nakamura}}, \bibinfo {author} {\bibfnamefont {H.~T.~H.}\
  \bibnamefont {Tashiro}},\ and\ \bibinfo {author} {\bibfnamefont {K.~T.~K.}\
  \bibnamefont {Toyoda}},\ }\bibfield  {title} {\bibinfo {title}
  {Micropatterning of quartz substrates by {Multi-WavelengthVacuum-Ultraviolet}
  laser ablation},\ }\href@noop {} {\bibfield  {journal} {\bibinfo  {journal}
  {Japanese Journal of Applied Physics}\ }\textbf {\bibinfo {volume} {32}},\
  \bibinfo {pages} {6185} (\bibinfo {year} {1993})}\BibitemShut {NoStop}%
\bibitem [{\citenamefont {Zhang}\ \emph {et~al.}(1997)\citenamefont {Zhang},
  \citenamefont {Sugioka}, \citenamefont {Wada}, \citenamefont {Tashiro},\ and\
  \citenamefont {Toyoda}}]{Zhang1997-of}%
  \BibitemOpen
  \bibfield  {author} {\bibinfo {author} {\bibfnamefont {J.}~\bibnamefont
  {Zhang}}, \bibinfo {author} {\bibfnamefont {K.}~\bibnamefont {Sugioka}},
  \bibinfo {author} {\bibfnamefont {S.}~\bibnamefont {Wada}}, \bibinfo {author}
  {\bibfnamefont {H.}~\bibnamefont {Tashiro}},\ and\ \bibinfo {author}
  {\bibfnamefont {K.}~\bibnamefont {Toyoda}},\ }\bibfield  {title} {\bibinfo
  {title} {Dual-beam ablation of fused quartz using 266 nm and {VUV} lasers
  with different delay-times},\ }\href@noop {} {\bibfield  {journal} {\bibinfo
  {journal} {Applied Physics A}\ }\textbf {\bibinfo {volume} {64}},\ \bibinfo
  {pages} {477} (\bibinfo {year} {1997})}\BibitemShut {NoStop}%
\bibitem [{\citenamefont {Zhang}\ \emph {et~al.}(2000)\citenamefont {Zhang},
  \citenamefont {Sugioka}, \citenamefont {Takahashi}, \citenamefont {Toyoda},\
  and\ \citenamefont {Midorikawa}}]{Zhang2000-ev}%
  \BibitemOpen
  \bibfield  {author} {\bibinfo {author} {\bibfnamefont {J.}~\bibnamefont
  {Zhang}}, \bibinfo {author} {\bibfnamefont {K.}~\bibnamefont {Sugioka}},
  \bibinfo {author} {\bibfnamefont {T.}~\bibnamefont {Takahashi}}, \bibinfo
  {author} {\bibfnamefont {K.}~\bibnamefont {Toyoda}},\ and\ \bibinfo {author}
  {\bibfnamefont {K.}~\bibnamefont {Midorikawa}},\ }\bibfield  {title}
  {\bibinfo {title} {Dual-beam ablation of fused silica by multiwavelength
  excitation process using {KrF} excimer and {F2} lasers},\ }\href@noop {}
  {\bibfield  {journal} {\bibinfo  {journal} {Applied Physics A}\ }\textbf
  {\bibinfo {volume} {71}},\ \bibinfo {pages} {23} (\bibinfo {year}
  {2000})}\BibitemShut {NoStop}%
\bibitem [{\citenamefont {Obata}\ \emph {et~al.}(2001)\citenamefont {Obata},
  \citenamefont {Sugioka}, \citenamefont {Akane}, \citenamefont {Aoki},
  \citenamefont {Toyoda},\ and\ \citenamefont {Midorikawa}}]{Obata2001-ey}%
  \BibitemOpen
  \bibfield  {author} {\bibinfo {author} {\bibfnamefont {K.}~\bibnamefont
  {Obata}}, \bibinfo {author} {\bibfnamefont {K.}~\bibnamefont {Sugioka}},
  \bibinfo {author} {\bibfnamefont {T.}~\bibnamefont {Akane}}, \bibinfo
  {author} {\bibfnamefont {N.}~\bibnamefont {Aoki}}, \bibinfo {author}
  {\bibfnamefont {K.}~\bibnamefont {Toyoda}},\ and\ \bibinfo {author}
  {\bibfnamefont {K.}~\bibnamefont {Midorikawa}},\ }\bibfield  {title}
  {\bibinfo {title} {Influence of laser fluence and irradiation timing of {F2}
  laser on ablation properties of fused silica in {F2-KrF} excimer laser
  multi-wavelength excitation process},\ }\href@noop {} {\bibfield  {journal}
  {\bibinfo  {journal} {Applied Physics A}\ }\textbf {\bibinfo {volume} {73}},\
  \bibinfo {pages} {755} (\bibinfo {year} {2001})}\BibitemShut {NoStop}%
\bibitem [{\citenamefont {Zoppel}\ \emph {et~al.}(2005)\citenamefont {Zoppel},
  \citenamefont {Merz}, \citenamefont {Zehetner},\ and\ \citenamefont
  {Reiderer}}]{Zoppel2005-En}%
  \BibitemOpen
  \bibfield  {author} {\bibinfo {author} {\bibfnamefont {S.}~\bibnamefont
  {Zoppel}}, \bibinfo {author} {\bibfnamefont {R.}~\bibnamefont {Merz}},
  \bibinfo {author} {\bibfnamefont {J.}~\bibnamefont {Zehetner}},\ and\
  \bibinfo {author} {\bibfnamefont {G.}~\bibnamefont {Reiderer}},\ }\bibfield
  {title} {\bibinfo {title} {Enhancement of laser ablation yield by two color
  excitation},\ }\href@noop {} {\bibfield  {journal} {\bibinfo  {journal}
  {Applied Physics A}\ }\textbf {\bibinfo {volume} {81}},\ \bibinfo {pages}
  {847} (\bibinfo {year} {2005})}\BibitemShut {NoStop}%
\bibitem [{\citenamefont {Zoppel}\ \emph {et~al.}(2007)\citenamefont {Zoppel},
  \citenamefont {Zehetner},\ and\ \citenamefont {Reider}}]{Zoppel2007-Tw}%
  \BibitemOpen
  \bibfield  {author} {\bibinfo {author} {\bibfnamefont {S.}~\bibnamefont
  {Zoppel}}, \bibinfo {author} {\bibfnamefont {J.}~\bibnamefont {Zehetner}},\
  and\ \bibinfo {author} {\bibfnamefont {G.~A.}\ \bibnamefont {Reider}},\
  }\bibfield  {title} {\bibinfo {title} {Two color laser ablation: Enhanced
  yield, improved machining},\ }\href@noop {} {\bibfield  {journal} {\bibinfo
  {journal} {Applied Surface Science}\ }\textbf {\bibinfo {volume} {253}},\
  \bibinfo {pages} {7692} (\bibinfo {year} {2007})}\BibitemShut {NoStop}%
\bibitem [{\citenamefont {Yu}\ \emph {et~al.}(2013)\citenamefont {Yu},
  \citenamefont {Bian}, \citenamefont {Chang}, \citenamefont {Corkum},\ and\
  \citenamefont {Lei}}]{Yu2013-aw}%
  \BibitemOpen
  \bibfield  {author} {\bibinfo {author} {\bibfnamefont {X.}~\bibnamefont
  {Yu}}, \bibinfo {author} {\bibfnamefont {Q.}~\bibnamefont {Bian}}, \bibinfo
  {author} {\bibfnamefont {Z.}~\bibnamefont {Chang}}, \bibinfo {author}
  {\bibfnamefont {P.~B.}\ \bibnamefont {Corkum}},\ and\ \bibinfo {author}
  {\bibfnamefont {S.}~\bibnamefont {Lei}},\ }\bibfield  {title} {\bibinfo
  {title} {Femtosecond laser nanomachining initiated by ultraviolet multiphoton
  ionization},\ }\href@noop {} {\bibfield  {journal} {\bibinfo  {journal}
  {Optics Express}\ }\textbf {\bibinfo {volume} {21}},\ \bibinfo {pages}
  {24185} (\bibinfo {year} {2013})}\BibitemShut {NoStop}%
\bibitem [{\citenamefont {Yu}\ \emph {et~al.}(2014)\citenamefont {Yu},
  \citenamefont {Chang}, \citenamefont {Corkum},\ and\ \citenamefont
  {Lei}}]{Yu2014-Fa}%
  \BibitemOpen
  \bibfield  {author} {\bibinfo {author} {\bibfnamefont {X.}~\bibnamefont
  {Yu}}, \bibinfo {author} {\bibfnamefont {Z.}~\bibnamefont {Chang}}, \bibinfo
  {author} {\bibfnamefont {P.~B.}\ \bibnamefont {Corkum}},\ and\ \bibinfo
  {author} {\bibfnamefont {S.}~\bibnamefont {Lei}},\ }\bibfield  {title}
  {\bibinfo {title} {Fabricating nanostructures on fused silica using
  femtosecond infrared pulses combined with sub-nanojoule ultraviolet pulses},\
  }\href@noop {} {\bibfield  {journal} {\bibinfo  {journal} {Optics Letters}\
  }\textbf {\bibinfo {volume} {39}},\ \bibinfo {pages} {19} (\bibinfo {year}
  {2014})}\BibitemShut {NoStop}%
\bibitem [{\citenamefont {Yang}\ \emph {et~al.}(2015)\citenamefont {Yang},
  \citenamefont {Lin}, \citenamefont {Zaytsev}, \citenamefont {Teng},
  \citenamefont {Her},\ and\ \citenamefont {Pan}}]{Yang2015-Fe}%
  \BibitemOpen
  \bibfield  {author} {\bibinfo {author} {\bibfnamefont {C.-S.}\ \bibnamefont
  {Yang}}, \bibinfo {author} {\bibfnamefont {C.-H.}\ \bibnamefont {Lin}},
  \bibinfo {author} {\bibfnamefont {A.}~\bibnamefont {Zaytsev}}, \bibinfo
  {author} {\bibfnamefont {K.-C.}\ \bibnamefont {Teng}}, \bibinfo {author}
  {\bibfnamefont {T.-H.}\ \bibnamefont {Her}},\ and\ \bibinfo {author}
  {\bibfnamefont {C.-L.}\ \bibnamefont {Pan}},\ }\bibfield  {title} {\bibinfo
  {title} {Femtosecond laser ablation of polymethylmethacrylate via dual-color
  synthesized waveform},\ }\href@noop {} {\bibfield  {journal} {\bibinfo
  {journal} {Applied Physics Letters}\ }\textbf {\bibinfo {volume} {106}},\
  \bibinfo {pages} {051902} (\bibinfo {year} {2015})}\BibitemShut {NoStop}%
\bibitem [{\citenamefont {Gedvilas}\ \emph {et~al.}(2017)\citenamefont
  {Gedvilas}, \citenamefont {Mik\v{s}y}, \citenamefont {Berzin\v{s}},
  \citenamefont {Stankevi\v{c}i},\ and\ \citenamefont
  {Ra\v{c}iukaitis}}]{Gedvilas2017-Mu}%
  \BibitemOpen
  \bibfield  {author} {\bibinfo {author} {\bibfnamefont {M.}~\bibnamefont
  {Gedvilas}}, \bibinfo {author} {\bibfnamefont {J.}~\bibnamefont {Mik\v{s}y}},
  \bibinfo {author} {\bibfnamefont {J.}~\bibnamefont {Berzin\v{s}}}, \bibinfo
  {author} {\bibfnamefont {V.}~\bibnamefont {Stankevi\v{c}i}},\ and\ \bibinfo
  {author} {\bibfnamefont {G.}~\bibnamefont {Ra\v{c}iukaitis}},\ }\bibfield
  {title} {\bibinfo {title} {Multi-photon absorption enhancement by
  dual-wavelength double-pulse laser irradiation for efficient dicing of
  sapphire wafers},\ }\href@noop {} {\bibfield  {journal} {\bibinfo  {journal}
  {Scientific reports}\ }\textbf {\bibinfo {volume} {7}},\ \bibinfo {pages}
  {5218} (\bibinfo {year} {2017})}\BibitemShut {NoStop}%
\bibitem [{\citenamefont {Keldysh}(1965)}]{Keldysh1965}%
  \BibitemOpen
  \bibfield  {author} {\bibinfo {author} {\bibfnamefont {L.~V.}\ \bibnamefont
  {Keldysh}},\ }\bibfield  {title} {\bibinfo {title} {Ionization in the field
  of a strong electromagnetic wave},\ }\href@noop {} {\bibfield  {journal}
  {\bibinfo  {journal} {Soviet Physics JETP}\ }\textbf {\bibinfo {volume}
  {20}},\ \bibinfo {pages} {1307} (\bibinfo {year} {1965})}\BibitemShut
  {NoStop}%
\bibitem [{\citenamefont {Shcheblanov}\ \emph {et~al.}(2017)\citenamefont
  {Shcheblanov}, \citenamefont {Povarnitsyn}, \citenamefont {Terekhin},
  \citenamefont {Guizard},\ and\ \citenamefont
  {Couairon}}]{Shcheblanov2017-em}%
  \BibitemOpen
  \bibfield  {author} {\bibinfo {author} {\bibfnamefont {N.~S.}\ \bibnamefont
  {Shcheblanov}}, \bibinfo {author} {\bibfnamefont {M.~E.}\ \bibnamefont
  {Povarnitsyn}}, \bibinfo {author} {\bibfnamefont {P.~N.}\ \bibnamefont
  {Terekhin}}, \bibinfo {author} {\bibfnamefont {S.}~\bibnamefont {Guizard}},\
  and\ \bibinfo {author} {\bibfnamefont {A.}~\bibnamefont {Couairon}},\
  }\bibfield  {title} {\bibinfo {title} {Nonlinear photoionization of
  transparent solids: A nonperturbative theory obeying selection rules},\
  }\href@noop {} {\bibfield  {journal} {\bibinfo  {journal} {Physical Review
  A}\ }\textbf {\bibinfo {volume} {96}},\ \bibinfo {pages} {063410} (\bibinfo
  {year} {2017})}\BibitemShut {NoStop}%
\bibitem [{\citenamefont {Otobe}\ \emph {et~al.}(2019)\citenamefont {Otobe},
  \citenamefont {Shinohara}, \citenamefont {Sato},\ and\ \citenamefont
  {Yabana}}]{Otobe2019-cv}%
  \BibitemOpen
  \bibfield  {author} {\bibinfo {author} {\bibfnamefont {T.}~\bibnamefont
  {Otobe}}, \bibinfo {author} {\bibfnamefont {Y.}~\bibnamefont {Shinohara}},
  \bibinfo {author} {\bibfnamefont {S.~A.}\ \bibnamefont {Sato}},\ and\
  \bibinfo {author} {\bibfnamefont {K.}~\bibnamefont {Yabana}},\ }\bibfield
  {title} {\bibinfo {title} {Theory for electron excitation in dielectrics
  under an intense linear and circularly polarized laser fields},\ }\href@noop
  {} {\bibfield  {journal} {\bibinfo  {journal} {Journal of the Physical
  Society of Japan}\ }\textbf {\bibinfo {volume} {88}},\ \bibinfo {pages}
  {024706} (\bibinfo {year} {2019})}\BibitemShut {NoStop}%
\bibitem [{\citenamefont {Bloembergen}(1974)}]{Bloembergen1974-hi}%
  \BibitemOpen
  \bibfield  {author} {\bibinfo {author} {\bibfnamefont {N.}~\bibnamefont
  {Bloembergen}},\ }\bibfield  {title} {\bibinfo {title} {Laser-induced
  electric breakdown in solids},\ }\href@noop {} {\bibfield  {journal}
  {\bibinfo  {journal} {IEEE Journal of Quantum Electronics}\ }\textbf
  {\bibinfo {volume} {10}},\ \bibinfo {pages} {375} (\bibinfo {year}
  {1974})}\BibitemShut {NoStop}%
\bibitem [{\citenamefont {Lagomarsino}\ \emph {et~al.}(2016)\citenamefont
  {Lagomarsino}, \citenamefont {Sciortino}, \citenamefont {Obreshkov},
  \citenamefont {Apostolova}, \citenamefont {Corsi}, \citenamefont {Bellini},
  \citenamefont {Berdermann},\ and\ \citenamefont
  {Schmidt}}]{Lagomarsino2016-by}%
  \BibitemOpen
  \bibfield  {author} {\bibinfo {author} {\bibfnamefont {S.}~\bibnamefont
  {Lagomarsino}}, \bibinfo {author} {\bibfnamefont {S.}~\bibnamefont
  {Sciortino}}, \bibinfo {author} {\bibfnamefont {B.}~\bibnamefont
  {Obreshkov}}, \bibinfo {author} {\bibfnamefont {T.}~\bibnamefont
  {Apostolova}}, \bibinfo {author} {\bibfnamefont {C.}~\bibnamefont {Corsi}},
  \bibinfo {author} {\bibfnamefont {M.}~\bibnamefont {Bellini}}, \bibinfo
  {author} {\bibfnamefont {E.}~\bibnamefont {Berdermann}},\ and\ \bibinfo
  {author} {\bibfnamefont {C.~J.}\ \bibnamefont {Schmidt}},\ }\bibfield
  {title} {\bibinfo {title} {Photoionization of monocrystalline {CVD} diamond
  irradiated with ultrashort intense laser pulse},\ }\href@noop {} {\bibfield
  {journal} {\bibinfo  {journal} {Physical Review B}\ }\textbf {\bibinfo
  {volume} {93}},\ \bibinfo {pages} {085128} (\bibinfo {year}
  {2016})}\BibitemShut {NoStop}%
\bibitem [{\citenamefont {Ikemachi}\ \emph {et~al.}(2017)\citenamefont
  {Ikemachi}, \citenamefont {Shinohara}, \citenamefont {Sato}, \citenamefont
  {Yumoto}, \citenamefont {Kuwata-Gonokami},\ and\ \citenamefont
  {Ishikawa}}]{PhysRevA.95.043416}%
  \BibitemOpen
  \bibfield  {author} {\bibinfo {author} {\bibfnamefont {T.}~\bibnamefont
  {Ikemachi}}, \bibinfo {author} {\bibfnamefont {Y.}~\bibnamefont {Shinohara}},
  \bibinfo {author} {\bibfnamefont {T.}~\bibnamefont {Sato}}, \bibinfo {author}
  {\bibfnamefont {J.}~\bibnamefont {Yumoto}}, \bibinfo {author} {\bibfnamefont
  {M.}~\bibnamefont {Kuwata-Gonokami}},\ and\ \bibinfo {author} {\bibfnamefont
  {K.~L.}\ \bibnamefont {Ishikawa}},\ }\bibfield  {title} {\bibinfo {title}
  {Trajectory analysis of high-order-harmonic generation from periodic
  crystals},\ }\href {https://doi.org/10.1103/PhysRevA.95.043416} {\bibfield
  {journal} {\bibinfo  {journal} {Physical Review A}\ }\textbf {\bibinfo
  {volume} {95}},\ \bibinfo {pages} {043416} (\bibinfo {year}
  {2017})}\BibitemShut {NoStop}%
\bibitem [{\citenamefont {Lindberg}\ and\ \citenamefont
  {Koch}(1988)}]{Lindberg1988-pr}%
  \BibitemOpen
  \bibfield  {author} {\bibinfo {author} {\bibfnamefont {M.}~\bibnamefont
  {Lindberg}}\ and\ \bibinfo {author} {\bibfnamefont {S.~W.}\ \bibnamefont
  {Koch}},\ }\bibfield  {title} {\bibinfo {title} {Effective bloch equations
  for semiconductors},\ }\href@noop {} {\bibfield  {journal} {\bibinfo
  {journal} {Physical Review B}\ }\textbf {\bibinfo {volume} {38}},\ \bibinfo
  {pages} {3342} (\bibinfo {year} {1988})}\BibitemShut {NoStop}%
\bibitem [{\citenamefont {Kaneshima}\ \emph {et~al.}(2018)\citenamefont
  {Kaneshima}, \citenamefont {Shinohara}, \citenamefont {Takeuchi},
  \citenamefont {Ishii}, \citenamefont {Imasaka}, \citenamefont {Kaji},
  \citenamefont {Ashihara}, \citenamefont {Ishikawa},\ and\ \citenamefont
  {Itatani}}]{DM}%
  \BibitemOpen
  \bibfield  {author} {\bibinfo {author} {\bibfnamefont {K.}~\bibnamefont
  {Kaneshima}}, \bibinfo {author} {\bibfnamefont {Y.}~\bibnamefont
  {Shinohara}}, \bibinfo {author} {\bibfnamefont {K.}~\bibnamefont {Takeuchi}},
  \bibinfo {author} {\bibfnamefont {N.}~\bibnamefont {Ishii}}, \bibinfo
  {author} {\bibfnamefont {K.}~\bibnamefont {Imasaka}}, \bibinfo {author}
  {\bibfnamefont {T.}~\bibnamefont {Kaji}}, \bibinfo {author} {\bibfnamefont
  {S.}~\bibnamefont {Ashihara}}, \bibinfo {author} {\bibfnamefont {K.~L.}\
  \bibnamefont {Ishikawa}},\ and\ \bibinfo {author} {\bibfnamefont
  {J.}~\bibnamefont {Itatani}},\ }\bibfield  {title} {\bibinfo {title}
  {Polarization-resolved study of high harmonics from bulk semiconductors},\
  }\href@noop {} {\bibfield  {journal} {\bibinfo  {journal} {Physical
  \red{R}eview \red{L}etters}\ }\textbf {\bibinfo {volume} {120}},\ \bibinfo
  {pages} {243903} (\bibinfo {year} {2018})}\BibitemShut {NoStop}%
\bibitem [{\citenamefont {Sanari}\ \emph {et~al.}(2020)\citenamefont {Sanari},
  \citenamefont {Hirori}, \citenamefont {Aharen}, \citenamefont {Tahara},
  \citenamefont {Shinohara}, \citenamefont {Ishikawa}, \citenamefont {Otobe},
  \citenamefont {Xia}, \citenamefont {Ishii}, \citenamefont {Itatani},
  \citenamefont {Sato},\ and\ \citenamefont {Kanemitsu}}]{PhysRevB.102.041125}%
  \BibitemOpen
  \bibfield  {author} {\bibinfo {author} {\bibfnamefont {Y.}~\bibnamefont
  {Sanari}}, \bibinfo {author} {\bibfnamefont {H.}~\bibnamefont {Hirori}},
  \bibinfo {author} {\bibfnamefont {T.}~\bibnamefont {Aharen}}, \bibinfo
  {author} {\bibfnamefont {H.}~\bibnamefont {Tahara}}, \bibinfo {author}
  {\bibfnamefont {Y.}~\bibnamefont {Shinohara}}, \bibinfo {author}
  {\bibfnamefont {K.~L.}\ \bibnamefont {Ishikawa}}, \bibinfo {author}
  {\bibfnamefont {T.}~\bibnamefont {Otobe}}, \bibinfo {author} {\bibfnamefont
  {P.}~\bibnamefont {Xia}}, \bibinfo {author} {\bibfnamefont {N.}~\bibnamefont
  {Ishii}}, \bibinfo {author} {\bibfnamefont {J.}~\bibnamefont {Itatani}},
  \bibinfo {author} {\bibfnamefont {S.~A.}\ \bibnamefont {Sato}},\ and\
  \bibinfo {author} {\bibfnamefont {Y.}~\bibnamefont {Kanemitsu}},\ }\bibfield
  {title} {\bibinfo {title} {Role of virtual band population for high harmonic
  generation in solids},\ }\href {https://doi.org/10.1103/PhysRevB.102.041125}
  {\bibfield  {journal} {\bibinfo  {journal} {Physical Review B}\ }\textbf
  {\bibinfo {volume} {102}},\ \bibinfo {pages} {041125} (\bibinfo {year}
  {2020})}\BibitemShut {NoStop}%
\bibitem [{\citenamefont {Hirori}\ \emph {et~al.}(2019)\citenamefont {Hirori},
  \citenamefont {Xia}, \citenamefont {Shinohara}, \citenamefont {Otobe},
  \citenamefont {Sanari}, \citenamefont {Tahara}, \citenamefont {Ishii},
  \citenamefont {Itatani}, \citenamefont {Ishikawa}, \citenamefont {Aharen},
  \citenamefont {Ozaki}, \citenamefont {Wakamiya},\ and\ \citenamefont
  {Kanemitsu}}]{HiroriAPL2019}%
  \BibitemOpen
  \bibfield  {author} {\bibinfo {author} {\bibfnamefont {H.}~\bibnamefont
  {Hirori}}, \bibinfo {author} {\bibfnamefont {P.}~\bibnamefont {Xia}},
  \bibinfo {author} {\bibfnamefont {Y.}~\bibnamefont {Shinohara}}, \bibinfo
  {author} {\bibfnamefont {T.}~\bibnamefont {Otobe}}, \bibinfo {author}
  {\bibfnamefont {Y.}~\bibnamefont {Sanari}}, \bibinfo {author} {\bibfnamefont
  {H.}~\bibnamefont {Tahara}}, \bibinfo {author} {\bibfnamefont
  {N.}~\bibnamefont {Ishii}}, \bibinfo {author} {\bibfnamefont
  {J.}~\bibnamefont {Itatani}}, \bibinfo {author} {\bibfnamefont {K.~L.}\
  \bibnamefont {Ishikawa}}, \bibinfo {author} {\bibfnamefont {T.}~\bibnamefont
  {Aharen}}, \bibinfo {author} {\bibfnamefont {M.}~\bibnamefont {Ozaki}},
  \bibinfo {author} {\bibfnamefont {A.}~\bibnamefont {Wakamiya}},\ and\
  \bibinfo {author} {\bibfnamefont {Y.}~\bibnamefont {Kanemitsu}},\ }\bibfield
  {title} {\bibinfo {title} {High-order harmonic generation from hybrid
  organic–inorganic perovskite thin films},\ }\href
  {https://doi.org/10.1063/1.5090935} {\bibfield  {journal} {\bibinfo
  {journal} {APL Materials}\ }\textbf {\bibinfo {volume} {7}},\ \bibinfo
  {pages} {041107} (\bibinfo {year} {2019})}\BibitemShut {NoStop}%
\bibitem [{\citenamefont {Yue}\ and\ \citenamefont
  {Gaarde}(2020)}]{Yue2020-rm}%
  \BibitemOpen
  \bibfield  {author} {\bibinfo {author} {\bibfnamefont {L.}~\bibnamefont
  {Yue}}\ and\ \bibinfo {author} {\bibfnamefont {M.~B.}\ \bibnamefont
  {Gaarde}},\ }\bibfield  {title} {\bibinfo {title} {Structure gauges and laser
  gauges for the semiconductor bloch equations in high-order harmonic
  generation in solids},\ }\href@noop {} {\bibfield  {journal} {\bibinfo
  {journal} {Phys. Rev. A}\ }\textbf {\bibinfo {volume} {101}},\ \bibinfo
  {pages} {053411} (\bibinfo {year} {2020})}\BibitemShut {NoStop}%
\bibitem [{\citenamefont {Otobe}\ \emph {et~al.}(2008)\citenamefont {Otobe},
  \citenamefont {Yamagiwa}, \citenamefont {Iwata}, \citenamefont {Yabana},
  \citenamefont {Nakatsukasa},\ and\ \citenamefont {Bertsch}}]{Otobe2008-yz}%
  \BibitemOpen
  \bibfield  {author} {\bibinfo {author} {\bibfnamefont {T.}~\bibnamefont
  {Otobe}}, \bibinfo {author} {\bibfnamefont {M.}~\bibnamefont {Yamagiwa}},
  \bibinfo {author} {\bibfnamefont {J.-I.}\ \bibnamefont {Iwata}}, \bibinfo
  {author} {\bibfnamefont {K.}~\bibnamefont {Yabana}}, \bibinfo {author}
  {\bibfnamefont {T.}~\bibnamefont {Nakatsukasa}},\ and\ \bibinfo {author}
  {\bibfnamefont {G.~F.}\ \bibnamefont {Bertsch}},\ }\bibfield  {title}
  {\bibinfo {title} {First-principles electron dynamics simulation for optical
  breakdown of dielectrics under an intense laser field},\ }\href@noop {}
  {\bibfield  {journal} {\bibinfo  {journal} {Physical Review B}\ }\textbf
  {\bibinfo {volume} {77}},\ \bibinfo {pages} {165104} (\bibinfo {year}
  {2008})}\BibitemShut {NoStop}%
\bibitem [{\citenamefont {Wachter}\ \emph {et~al.}(2014)\citenamefont
  {Wachter}, \citenamefont {Lemell}, \citenamefont {Burgd{\"o}rfer},
  \citenamefont {Sato}, \citenamefont {Tong},\ and\ \citenamefont
  {Yabana}}]{Wachter2014-pg}%
  \BibitemOpen
  \bibfield  {author} {\bibinfo {author} {\bibfnamefont {G.}~\bibnamefont
  {Wachter}}, \bibinfo {author} {\bibfnamefont {C.}~\bibnamefont {Lemell}},
  \bibinfo {author} {\bibfnamefont {J.}~\bibnamefont {Burgd{\"o}rfer}},
  \bibinfo {author} {\bibfnamefont {S.~A.}\ \bibnamefont {Sato}}, \bibinfo
  {author} {\bibfnamefont {X.-M.}\ \bibnamefont {Tong}},\ and\ \bibinfo
  {author} {\bibfnamefont {K.}~\bibnamefont {Yabana}},\ }\bibfield  {title}
  {\bibinfo {title} {Ab initio simulation of electrical currents induced by
  ultrafast laser excitation of dielectric materials},\ }\href@noop {}
  {\bibfield  {journal} {\bibinfo  {journal} {Physical Review Letters}\
  }\textbf {\bibinfo {volume} {113}},\ \bibinfo {pages} {087401} (\bibinfo
  {year} {2014})}\BibitemShut {NoStop}%
\bibitem [{\citenamefont {Yamada}\ and\ \citenamefont
  {Yabana}(2019)}]{Yamada2019-ls}%
  \BibitemOpen
  \bibfield  {author} {\bibinfo {author} {\bibfnamefont {A.}~\bibnamefont
  {Yamada}}\ and\ \bibinfo {author} {\bibfnamefont {K.}~\bibnamefont
  {Yabana}},\ }\bibfield  {title} {\bibinfo {title} {Energy transfer from
  intense laser pulse to dielectrics in time-dependent density functional
  theory},\ }\href@noop {} {\bibfield  {journal} {\bibinfo  {journal} {The E
  uropean Physical Journal D}\ }\textbf {\bibinfo {volume} {73}},\ \bibinfo
  {pages} {87} (\bibinfo {year} {2019})}\BibitemShut {NoStop}%
\bibitem [{\citenamefont {Yamada}\ and\ \citenamefont
  {Yabana}(2020)}]{Yamada2020-qt}%
  \BibitemOpen
  \bibfield  {author} {\bibinfo {author} {\bibfnamefont {S.}~\bibnamefont
  {Yamada}}\ and\ \bibinfo {author} {\bibfnamefont {K.}~\bibnamefont
  {Yabana}},\ }\bibfield  {title} {\bibinfo {title} {Symmetry properties of
  attosecond transient absorption spectroscopy in crystalline dielectrics},\
  }\href@noop {} {\bibfield  {journal} {\bibinfo  {journal} {Physical Review
  B}\ }\textbf {\bibinfo {volume} {101}},\ \bibinfo {pages} {165128} (\bibinfo
  {year} {2020})}\BibitemShut {NoStop}%
\bibitem [{\citenamefont {Hui}\ \emph {et~al.}(2021)\citenamefont {Hui},
  \citenamefont {Alqattan}, \citenamefont {Yamada}, \citenamefont {Pervak},
  \citenamefont {Yabana},\ and\ \citenamefont {Hassan}}]{Hui2021-cy}%
  \BibitemOpen
  \bibfield  {author} {\bibinfo {author} {\bibfnamefont {D.}~\bibnamefont
  {Hui}}, \bibinfo {author} {\bibfnamefont {H.}~\bibnamefont {Alqattan}},
  \bibinfo {author} {\bibfnamefont {S.}~\bibnamefont {Yamada}}, \bibinfo
  {author} {\bibfnamefont {V.}~\bibnamefont {Pervak}}, \bibinfo {author}
  {\bibfnamefont {K.}~\bibnamefont {Yabana}},\ and\ \bibinfo {author}
  {\bibfnamefont {M.~T.}\ \bibnamefont {Hassan}},\ }\bibfield  {title}
  {\bibinfo {title} {Attosecond electron motion control in dielectric},\
  }\href@noop {} {\bibfield  {journal} {\bibinfo  {journal} {Nature Photonics}\
  }\textbf {\bibinfo {volume} {16}},\ \bibinfo {pages} {33} (\bibinfo {year}
  {2021})}\BibitemShut {NoStop}%
\bibitem [{\citenamefont {de~Alaiza~Mart\'{i}nez}\ \emph
  {et~al.}(2021)\citenamefont {de~Alaiza~Mart\'{i}nez}, \citenamefont
  {Smetanina}, \citenamefont {Thiele}, \citenamefont {Chimier},\ and\
  \citenamefont {Duchateau}}]{Martinez2021-Mo}%
  \BibitemOpen
  \bibfield  {author} {\bibinfo {author} {\bibfnamefont {P.~G.}\ \bibnamefont
  {de~Alaiza~Mart\'{i}nez}}, \bibinfo {author} {\bibfnamefont {E.}~\bibnamefont
  {Smetanina}}, \bibinfo {author} {\bibfnamefont {I.}~\bibnamefont {Thiele}},
  \bibinfo {author} {\bibfnamefont {B.}~\bibnamefont {Chimier}},\ and\ \bibinfo
  {author} {\bibfnamefont {G.}~\bibnamefont {Duchateau}},\ }\bibfield  {title}
  {\bibinfo {title} {Modeling the time-dependent electron dynamics in
  dielectric materials induced by two-color femtosecond laser pulses:
  Applications to material modifications},\ }\href@noop {} {\bibfield
  {journal} {\bibinfo  {journal} {Physical Review A}\ }\textbf {\bibinfo
  {volume} {103}},\ \bibinfo {pages} {033107} (\bibinfo {year}
  {2021})}\BibitemShut {NoStop}%
\bibitem [{\citenamefont {Duchateau}\ \emph {et~al.}(2022)\citenamefont
  {Duchateau}, \citenamefont {Yamada},\ and\ \citenamefont
  {Yabana}}]{Duchateau2022-ba}%
  \BibitemOpen
  \bibfield  {author} {\bibinfo {author} {\bibfnamefont {G.}~\bibnamefont
  {Duchateau}}, \bibinfo {author} {\bibfnamefont {A.}~\bibnamefont {Yamada}},\
  and\ \bibinfo {author} {\bibfnamefont {K.}~\bibnamefont {Yabana}},\
  }\bibfield  {title} {\bibinfo {title} {Electron dynamics in $\alpha$-quartz
  induced by two-color 10-femtosecond laser pulses},\ }\href@noop {} {\bibfield
   {journal} {\bibinfo  {journal} {Physical Review B}\ }\textbf {\bibinfo
  {volume} {105}},\ \bibinfo {pages} {165128} (\bibinfo {year}
  {2022})}\BibitemShut {NoStop}%
\bibitem [{\citenamefont {Golde}\ \emph {et~al.}(2008)\citenamefont {Golde},
  \citenamefont {Meier},\ and\ \citenamefont {Koch}}]{Golde2008-yi}%
  \BibitemOpen
  \bibfield  {author} {\bibinfo {author} {\bibfnamefont {D.}~\bibnamefont
  {Golde}}, \bibinfo {author} {\bibfnamefont {T.}~\bibnamefont {Meier}},\ and\
  \bibinfo {author} {\bibfnamefont {S.~W.}\ \bibnamefont {Koch}},\ }\bibfield
  {title} {\bibinfo {title} {High harmonics generated in semiconductor
  nanostructures by the coupled dynamics of optical inter- and intraband
  excitations},\ }\href@noop {} {\bibfield  {journal} {\bibinfo  {journal}
  {Physical Review B}\ }\textbf {\bibinfo {volume} {77}},\ \bibinfo {pages}
  {075330} (\bibinfo {year} {2008})}\BibitemShut {NoStop}%
\bibitem [{\citenamefont {Ghimire}\ \emph {et~al.}(2010)\citenamefont
  {Ghimire}, \citenamefont {DiChiara}, \citenamefont {Sistrunk}, \citenamefont
  {Agostini}, \citenamefont {DiMauro},\ and\ \citenamefont
  {Reis}}]{Ghimire2010-sg}%
  \BibitemOpen
  \bibfield  {author} {\bibinfo {author} {\bibfnamefont {S.}~\bibnamefont
  {Ghimire}}, \bibinfo {author} {\bibfnamefont {A.~D.}\ \bibnamefont
  {DiChiara}}, \bibinfo {author} {\bibfnamefont {E.}~\bibnamefont {Sistrunk}},
  \bibinfo {author} {\bibfnamefont {P.}~\bibnamefont {Agostini}}, \bibinfo
  {author} {\bibfnamefont {L.~F.}\ \bibnamefont {DiMauro}},\ and\ \bibinfo
  {author} {\bibfnamefont {D.~A.}\ \bibnamefont {Reis}},\ }\bibfield  {title}
  {\bibinfo {title} {Observation of high-order harmonic generation in a bulk
  crystal},\ }\href@noop {} {\bibfield  {journal} {\bibinfo  {journal} {Nat.
  Phys.}\ }\textbf {\bibinfo {volume} {7}},\ \bibinfo {pages} {138} (\bibinfo
  {year} {2010})}\BibitemShut {NoStop}%
\bibitem [{\citenamefont {McDonald}\ \emph {et~al.}(2015)\citenamefont
  {McDonald}, \citenamefont {Vampa}, \citenamefont {Orlando1}, \citenamefont
  {Corkum},\ and\ \citenamefont {Brabec}}]{McDonald2015}%
  \BibitemOpen
  \bibfield  {author} {\bibinfo {author} {\bibfnamefont {C.~R.}\ \bibnamefont
  {McDonald}}, \bibinfo {author} {\bibfnamefont {G.}~\bibnamefont {Vampa}},
  \bibinfo {author} {\bibfnamefont {G.}~\bibnamefont {Orlando1}}, \bibinfo
  {author} {\bibfnamefont {P.~B.}\ \bibnamefont {Corkum}},\ and\ \bibinfo
  {author} {\bibfnamefont {T.}~\bibnamefont {Brabec}},\ }\bibfield  {title}
  {\bibinfo {title} {Theory of high-harmonic generation in solids},\
  }\href@noop {} {\bibfield  {journal} {\bibinfo  {journal} {Journal of
  Physics: Conference Series}\ }\textbf {\bibinfo {volume} {594}},\ \bibinfo
  {pages} {012021} (\bibinfo {year} {2015})}\BibitemShut {NoStop}%
\bibitem [{\citenamefont {Wismer}\ \emph {et~al.}(2016)\citenamefont {Wismer},
  \citenamefont {Kruchinin}, \citenamefont {Ciappina}, \citenamefont
  {Stockman},\ and\ \citenamefont {Yakovlev}}]{Wismer2016-hi}%
  \BibitemOpen
  \bibfield  {author} {\bibinfo {author} {\bibfnamefont {M.~S.}\ \bibnamefont
  {Wismer}}, \bibinfo {author} {\bibfnamefont {S.~Y.}\ \bibnamefont
  {Kruchinin}}, \bibinfo {author} {\bibfnamefont {M.}~\bibnamefont {Ciappina}},
  \bibinfo {author} {\bibfnamefont {M.~I.}\ \bibnamefont {Stockman}},\ and\
  \bibinfo {author} {\bibfnamefont {V.~S.}\ \bibnamefont {Yakovlev}},\
  }\bibfield  {title} {\bibinfo {title} {{Strong-Field} resonant dynamics in
  semiconductors},\ }\href@noop {} {\bibfield  {journal} {\bibinfo  {journal}
  {Physical Review Letters}\ }\textbf {\bibinfo {volume} {116}},\ \bibinfo
  {pages} {197401} (\bibinfo {year} {2016})}\BibitemShut {NoStop}%
\bibitem [{\citenamefont {Wang}\ \emph {et~al.}(2018)\citenamefont {Wang},
  \citenamefont {Xu}, \citenamefont {Huang},\ and\ \citenamefont
  {Bian}}]{Wang2018-cd}%
  \BibitemOpen
  \bibfield  {author} {\bibinfo {author} {\bibfnamefont {X.-Q.}\ \bibnamefont
  {Wang}}, \bibinfo {author} {\bibfnamefont {Y.}~\bibnamefont {Xu}}, \bibinfo
  {author} {\bibfnamefont {X.-H.}\ \bibnamefont {Huang}},\ and\ \bibinfo
  {author} {\bibfnamefont {X.-B.}\ \bibnamefont {Bian}},\ }\bibfield  {title}
  {\bibinfo {title} {Interference between inter- and intraband currents in
  high-order harmonic generation in solids},\ }\href@noop {} {\bibfield
  {journal} {\bibinfo  {journal} {Physical Review A}\ }\textbf {\bibinfo
  {volume} {98}},\ \bibinfo {pages} {023427} (\bibinfo {year}
  {2018})}\BibitemShut {NoStop}%
\bibitem [{\citenamefont {Sato}\ \emph {et~al.}(2018)\citenamefont {Sato},
  \citenamefont {Lucchini}, \citenamefont {Volkov}, \citenamefont {Schlaepfer},
  \citenamefont {Gallmann}, \citenamefont {Keller},\ and\ \citenamefont
  {Rubio}}]{Sato2018-Ro}%
  \BibitemOpen
  \bibfield  {author} {\bibinfo {author} {\bibfnamefont {S.~A.}\ \bibnamefont
  {Sato}}, \bibinfo {author} {\bibfnamefont {M.}~\bibnamefont {Lucchini}},
  \bibinfo {author} {\bibfnamefont {M.}~\bibnamefont {Volkov}}, \bibinfo
  {author} {\bibfnamefont {F.}~\bibnamefont {Schlaepfer}}, \bibinfo {author}
  {\bibfnamefont {L.}~\bibnamefont {Gallmann}}, \bibinfo {author}
  {\bibfnamefont {U.}~\bibnamefont {Keller}},\ and\ \bibinfo {author}
  {\bibfnamefont {A.}~\bibnamefont {Rubio}},\ }\bibfield  {title} {\bibinfo
  {title} {Role of intraband transitions in photocarrier generation},\
  }\href@noop {} {\bibfield  {journal} {\bibinfo  {journal} {Physical Review
  B}\ }\textbf {\bibinfo {volume} {98}},\ \bibinfo {pages} {035202} (\bibinfo
  {year} {2018})}\BibitemShut {NoStop}%
\bibitem [{\citenamefont {Schlaepfer}\ \emph {et~al.}(2018)\citenamefont
  {Schlaepfer}, \citenamefont {Lucchini}, \citenamefont {Sato}, \citenamefont
  {Volkov}, \citenamefont {Kasmi}, \citenamefont {Hartmann}, \citenamefont
  {Rubio}, \citenamefont {Gallmann},\ and\ \citenamefont
  {Keller}}]{Schlaepfer2018-At}%
  \BibitemOpen
  \bibfield  {author} {\bibinfo {author} {\bibfnamefont {F.}~\bibnamefont
  {Schlaepfer}}, \bibinfo {author} {\bibfnamefont {M.}~\bibnamefont
  {Lucchini}}, \bibinfo {author} {\bibfnamefont {S.~A.}\ \bibnamefont {Sato}},
  \bibinfo {author} {\bibfnamefont {M.}~\bibnamefont {Volkov}}, \bibinfo
  {author} {\bibfnamefont {L.}~\bibnamefont {Kasmi}}, \bibinfo {author}
  {\bibfnamefont {N.}~\bibnamefont {Hartmann}}, \bibinfo {author}
  {\bibfnamefont {A.}~\bibnamefont {Rubio}}, \bibinfo {author} {\bibfnamefont
  {L.}~\bibnamefont {Gallmann}},\ and\ \bibinfo {author} {\bibfnamefont
  {U.}~\bibnamefont {Keller}},\ }\bibfield  {title} {\bibinfo {title}
  {Attosecond optical-field-enhanced carrier injection into the gaas conduction
  band},\ }\href@noop {} {\bibfield  {journal} {\bibinfo  {journal} {Nature
  Physics}\ }\textbf {\bibinfo {volume} {14}},\ \bibinfo {pages} {560}
  (\bibinfo {year} {2018})}\BibitemShut {NoStop}%
\bibitem [{\citenamefont {Song}\ \emph {et~al.}(2020)\citenamefont {Song},
  \citenamefont {Yang}, \citenamefont {Zuo}, \citenamefont {Meier},\ and\
  \citenamefont {Yang}}]{Song2020-hk}%
  \BibitemOpen
  \bibfield  {author} {\bibinfo {author} {\bibfnamefont {X.}~\bibnamefont
  {Song}}, \bibinfo {author} {\bibfnamefont {S.}~\bibnamefont {Yang}}, \bibinfo
  {author} {\bibfnamefont {R.}~\bibnamefont {Zuo}}, \bibinfo {author}
  {\bibfnamefont {T.}~\bibnamefont {Meier}},\ and\ \bibinfo {author}
  {\bibfnamefont {W.}~\bibnamefont {Yang}},\ }\bibfield  {title} {\bibinfo
  {title} {Enhanced high-order harmonic generation in semiconductors by
  excitation with multicolor pulses},\ }\href@noop {} {\bibfield  {journal}
  {\bibinfo  {journal} {Physical Review A}\ }\textbf {\bibinfo {volume}
  {101}},\ \bibinfo {pages} {033410} (\bibinfo {year} {2020})}\BibitemShut
  {NoStop}%
\bibitem [{\citenamefont {Koz\'ak}\ \emph {et~al.}(2020)\citenamefont
  {Koz\'ak}, \citenamefont {Mart\'inek}, \citenamefont {Otobe}, \citenamefont
  {Troj\'anek},\ and\ \citenamefont {Mal\'y}}]{Kozak2020-Ob}%
  \BibitemOpen
  \bibfield  {author} {\bibinfo {author} {\bibfnamefont {M.}~\bibnamefont
  {Koz\'ak}}, \bibinfo {author} {\bibfnamefont {M.}~\bibnamefont {Mart\'inek}},
  \bibinfo {author} {\bibfnamefont {T.}~\bibnamefont {Otobe}}, \bibinfo
  {author} {\bibfnamefont {F.}~\bibnamefont {Troj\'anek}},\ and\ \bibinfo
  {author} {\bibfnamefont {P.}~\bibnamefont {Mal\'y}},\ }\bibfield  {title}
  {\bibinfo {title} {Observation of ultrafast impact ionization in diamond
  driven by mid-infrared femtosecond pulses},\ }\href@noop {} {\bibfield
  {journal} {\bibinfo  {journal} {Journal of Applied Physics}\ }\textbf
  {\bibinfo {volume} {128}},\ \bibinfo {pages} {015701} (\bibinfo {year}
  {2020})}\BibitemShut {NoStop}%
\bibitem [{\citenamefont {Miyamoto}(2021)}]{Miyamoto2021-Di}%
  \BibitemOpen
  \bibfield  {author} {\bibinfo {author} {\bibfnamefont {Y.}~\bibnamefont
  {Miyamoto}},\ }\bibfield  {title} {\bibinfo {title} {Direct treatment of
  interaction between laser-field and electrons for simulating laser processing
  of metals},\ }\href@noop {} {\bibfield  {journal} {\bibinfo  {journal}
  {Scientific Reports}\ }\textbf {\bibinfo {volume} {11}},\ \bibinfo {pages}
  {14626} (\bibinfo {year} {2021})}\BibitemShut {NoStop}%
\bibitem [{\citenamefont {Otobe}(2016)}]{Otobe2016-Hi}%
  \BibitemOpen
  \bibfield  {author} {\bibinfo {author} {\bibfnamefont {T.}~\bibnamefont
  {Otobe}},\ }\bibfield  {title} {\bibinfo {title} {High-harmonic generation in
  $\alpha$-quartz by electron-hole recombination},\ }\href@noop {} {\bibfield
  {journal} {\bibinfo  {journal} {Physical Review B}\ }\textbf {\bibinfo
  {volume} {94}},\ \bibinfo {pages} {235152} (\bibinfo {year}
  {2016})}\BibitemShut {NoStop}%
\bibitem [{\citenamefont {Tancogne-Dejean}\ \emph {et~al.}(2017)\citenamefont
  {Tancogne-Dejean}, \citenamefont {M\"{u}cke}, \citenamefont {K\"{a}rtner},\
  and\ \citenamefont {Rubio}}]{Nicolas2017-Im}%
  \BibitemOpen
  \bibfield  {author} {\bibinfo {author} {\bibfnamefont {N.}~\bibnamefont
  {Tancogne-Dejean}}, \bibinfo {author} {\bibfnamefont {O.~D.}\ \bibnamefont
  {M\"{u}cke}}, \bibinfo {author} {\bibfnamefont {F.~X.}\ \bibnamefont
  {K\"{a}rtner}},\ and\ \bibinfo {author} {\bibfnamefont {A.}~\bibnamefont
  {Rubio}},\ }\bibfield  {title} {\bibinfo {title} {Impact of the electronic
  band structure in high-harmonic generation spectra of solids},\ }\href@noop
  {} {\bibfield  {journal} {\bibinfo  {journal} {Physical Review Letters}\
  }\textbf {\bibinfo {volume} {118}},\ \bibinfo {pages} {087403} (\bibinfo
  {year} {2017})}\BibitemShut {NoStop}%
\bibitem [{\citenamefont {Tancogne-Dejean}\ \emph {et~al.}(2018)\citenamefont
  {Tancogne-Dejean}, \citenamefont {Sentef},\ and\ \citenamefont
  {Rubio}}]{Nicolas2018-Ul}%
  \BibitemOpen
  \bibfield  {author} {\bibinfo {author} {\bibfnamefont {N.}~\bibnamefont
  {Tancogne-Dejean}}, \bibinfo {author} {\bibfnamefont {M.~A.}\ \bibnamefont
  {Sentef}},\ and\ \bibinfo {author} {\bibfnamefont {A.}~\bibnamefont
  {Rubio}},\ }\bibfield  {title} {\bibinfo {title} {Ultrafast modification of
  hubbard u in a strongly correlated material: Ab initio high-harmonic
  generation in nio},\ }\href@noop {} {\bibfield  {journal} {\bibinfo
  {journal} {Physical Review Letters}\ }\textbf {\bibinfo {volume} {121}},\
  \bibinfo {pages} {097402} (\bibinfo {year} {2018})}\BibitemShut {NoStop}%
\bibitem [{\citenamefont {Tancogne-Dejean}\ and\ \citenamefont
  {Rubio}(2018)}]{Nicolas2018-At}%
  \BibitemOpen
  \bibfield  {author} {\bibinfo {author} {\bibfnamefont {N.}~\bibnamefont
  {Tancogne-Dejean}}\ and\ \bibinfo {author} {\bibfnamefont {A.}~\bibnamefont
  {Rubio}},\ }\bibfield  {title} {\bibinfo {title} {Atomic-like high-harmonic
  generation from two-dimensional materials},\ }\href@noop {} {\bibfield
  {journal} {\bibinfo  {journal} {Science Advances}\ }\textbf {\bibinfo
  {volume} {4}},\ \bibinfo {pages} {eaao5207} (\bibinfo {year}
  {2018})}\BibitemShut {NoStop}%
\bibitem [{\citenamefont {Floss}\ \emph {et~al.}(2019)\citenamefont {Floss},
  \citenamefont {Lemell}, \citenamefont {Yabana},\ and\ \citenamefont
  {Burgd{\"o}rfer}}]{Floss2019-In}%
  \BibitemOpen
  \bibfield  {author} {\bibinfo {author} {\bibfnamefont {I.}~\bibnamefont
  {Floss}}, \bibinfo {author} {\bibfnamefont {C.}~\bibnamefont {Lemell}},
  \bibinfo {author} {\bibfnamefont {K.}~\bibnamefont {Yabana}},\ and\ \bibinfo
  {author} {\bibfnamefont {J.}~\bibnamefont {Burgd{\"o}rfer}},\ }\bibfield
  {title} {\bibinfo {title} {Incorporating decoherence into solid-state
  time-dependent density functional theory},\ }\href@noop {} {\bibfield
  {journal} {\bibinfo  {journal} {Physical Review B}\ }\textbf {\bibinfo
  {volume} {99}},\ \bibinfo {pages} {224301} (\bibinfo {year}
  {2019})}\BibitemShut {NoStop}%
\bibitem [{\citenamefont {Yamada}\ and\ \citenamefont
  {Yabana}(2021)}]{Yamada2021-De}%
  \BibitemOpen
  \bibfield  {author} {\bibinfo {author} {\bibfnamefont {S.}~\bibnamefont
  {Yamada}}\ and\ \bibinfo {author} {\bibfnamefont {K.}~\bibnamefont
  {Yabana}},\ }\bibfield  {title} {\bibinfo {title} {Determining the optimum
  thickness for high harmonic generation from nanoscale thin films: An ab
  initio computational study},\ }\href@noop {} {\bibfield  {journal} {\bibinfo
  {journal} {Physical Review B}\ }\textbf {\bibinfo {volume} {103}},\ \bibinfo
  {pages} {155426} (\bibinfo {year} {2021})}\BibitemShut {NoStop}%
\bibitem [{\citenamefont {Noda}\ \emph {et~al.}(2019)\citenamefont {Noda},
  \citenamefont {Sato}, \citenamefont {Hirokawa}, \citenamefont {Uemoto},
  \citenamefont {Takeuchi}, \citenamefont {Yamada}, \citenamefont {Yamada},
  \citenamefont {Shinohara}, \citenamefont {Yamaguchi}, \citenamefont {Iida},
  \citenamefont {Floss}, \citenamefont {Otobe}, \citenamefont {Lee},
  \citenamefont {Ishimura}, \citenamefont {Boku}, \citenamefont {Bertsch},
  \citenamefont {Nobusada},\ and\ \citenamefont {Yabana}}]{salmon}%
  \BibitemOpen
  \bibfield  {author} {\bibinfo {author} {\bibfnamefont {M.}~\bibnamefont
  {Noda}}, \bibinfo {author} {\bibfnamefont {S.~A.}\ \bibnamefont {Sato}},
  \bibinfo {author} {\bibfnamefont {Y.}~\bibnamefont {Hirokawa}}, \bibinfo
  {author} {\bibfnamefont {M.}~\bibnamefont {Uemoto}}, \bibinfo {author}
  {\bibfnamefont {T.}~\bibnamefont {Takeuchi}}, \bibinfo {author}
  {\bibfnamefont {S.}~\bibnamefont {Yamada}}, \bibinfo {author} {\bibfnamefont
  {A.}~\bibnamefont {Yamada}}, \bibinfo {author} {\bibfnamefont
  {Y.}~\bibnamefont {Shinohara}}, \bibinfo {author} {\bibfnamefont
  {M.}~\bibnamefont {Yamaguchi}}, \bibinfo {author} {\bibfnamefont
  {K.}~\bibnamefont {Iida}}, \bibinfo {author} {\bibfnamefont {I.}~\bibnamefont
  {Floss}}, \bibinfo {author} {\bibfnamefont {T.}~\bibnamefont {Otobe}},
  \bibinfo {author} {\bibfnamefont {K.-M.}\ \bibnamefont {Lee}}, \bibinfo
  {author} {\bibfnamefont {K.}~\bibnamefont {Ishimura}}, \bibinfo {author}
  {\bibfnamefont {T.}~\bibnamefont {Boku}}, \bibinfo {author} {\bibfnamefont
  {G.~F.}\ \bibnamefont {Bertsch}}, \bibinfo {author} {\bibfnamefont
  {K.}~\bibnamefont {Nobusada}},\ and\ \bibinfo {author} {\bibfnamefont
  {K.}~\bibnamefont {Yabana}},\ }\bibfield  {title} {\bibinfo {title}
  {{SALMON}: Scalable ab-initio light--matter simulator for optics and
  nanoscience},\ }\href@noop {} {\bibfield  {journal} {\bibinfo  {journal}
  {Computer Physics Communications}\ }\textbf {\bibinfo {volume} {235}},\
  \bibinfo {pages} {356} (\bibinfo {year} {2019})}\BibitemShut {NoStop}%
\bibitem [{\citenamefont {Runge}\ and\ \citenamefont {Gross}(1984)}]{TDDFT}%
  \BibitemOpen
  \bibfield  {author} {\bibinfo {author} {\bibfnamefont {E.}~\bibnamefont
  {Runge}}\ and\ \bibinfo {author} {\bibfnamefont {E.~K.}\ \bibnamefont
  {Gross}},\ }\bibfield  {title} {\bibinfo {title} {Density-functional theory
  for time-dependent systems},\ }\href@noop {} {\bibfield  {journal} {\bibinfo
  {journal} {Physical Review Letters}\ }\textbf {\bibinfo {volume} {52}},\
  \bibinfo {pages} {997} (\bibinfo {year} {1984})}\BibitemShut {NoStop}%
\bibitem [{\citenamefont {Fuchs}\ and\ \citenamefont {Scheffler}(1999)}]{FHI}%
  \BibitemOpen
  \bibfield  {author} {\bibinfo {author} {\bibfnamefont {M.}~\bibnamefont
  {Fuchs}}\ and\ \bibinfo {author} {\bibfnamefont {M.}~\bibnamefont
  {Scheffler}},\ }\bibfield  {title} {\bibinfo {title} {Ab initio
  pseudopotentials for electronic structure calculations of poly-atomic systems
  using density-functional theory},\ }\href@noop {} {\bibfield  {journal}
  {\bibinfo  {journal} {Computer Physics Communications}\ }\textbf {\bibinfo
  {volume} {119}},\ \bibinfo {pages} {67} (\bibinfo {year} {1999})}\BibitemShut
  {NoStop}%
\bibitem [{\citenamefont {Tran}\ and\ \citenamefont {Blaha}(2009)}]{TBmBJ}%
  \BibitemOpen
  \bibfield  {author} {\bibinfo {author} {\bibfnamefont {F.}~\bibnamefont
  {Tran}}\ and\ \bibinfo {author} {\bibfnamefont {P.}~\bibnamefont {Blaha}},\
  }\bibfield  {title} {\bibinfo {title} {Accurate band gaps of semiconductors
  and insulators with a semilocal exchange-correlation potential},\ }\href@noop
  {} {\bibfield  {journal} {\bibinfo  {journal} {Physical Review Letters}\
  }\textbf {\bibinfo {volume} {102}},\ \bibinfo {pages} {226401} (\bibinfo
  {year} {2009})}\BibitemShut {NoStop}%
\bibitem [{\citenamefont {Sato}\ \emph {et~al.}(2015)\citenamefont {Sato},
  \citenamefont {Taniguchi}, \citenamefont {Shinohara},\ and\ \citenamefont
  {Yabana}}]{Sato2015-No}%
  \BibitemOpen
  \bibfield  {author} {\bibinfo {author} {\bibfnamefont {S.~A.}\ \bibnamefont
  {Sato}}, \bibinfo {author} {\bibfnamefont {Y.}~\bibnamefont {Taniguchi}},
  \bibinfo {author} {\bibfnamefont {Y.}~\bibnamefont {Shinohara}},\ and\
  \bibinfo {author} {\bibfnamefont {K.}~\bibnamefont {Yabana}},\ }\bibfield
  {title} {\bibinfo {title} {Nonlinear electronic excitations in crystalline
  solids using meta-generalized gradient approximation and hybrid functional in
  time-dependent density functional theory},\ }\href@noop {} {\bibfield
  {journal} {\bibinfo  {journal} {Journal of Chemical Physics}\ }\textbf
  {\bibinfo {volume} {143}},\ \bibinfo {pages} {224116} (\bibinfo {year}
  {2015})}\BibitemShut {NoStop}%
\bibitem [{\citenamefont {Massa}\ \emph {et~al.}(2009)\citenamefont {Massa},
  \citenamefont {Mana}, \citenamefont {Kuetgens},\ and\ \citenamefont
  {Ferroglio}}]{Massa2009-hp}%
  \BibitemOpen
  \bibfield  {author} {\bibinfo {author} {\bibfnamefont {E.}~\bibnamefont
  {Massa}}, \bibinfo {author} {\bibfnamefont {G.}~\bibnamefont {Mana}},
  \bibinfo {author} {\bibfnamefont {U.}~\bibnamefont {Kuetgens}},\ and\
  \bibinfo {author} {\bibfnamefont {L.}~\bibnamefont {Ferroglio}},\ }\bibfield
  {title} {\bibinfo {title} {Measurement of the lattice parameter of a silicon
  crystal},\ }\href@noop {} {\bibfield  {journal} {\bibinfo  {journal} {New
  Journal of Physics}\ }\textbf {\bibinfo {volume} {11}},\ \bibinfo {pages}
  {053013} (\bibinfo {year} {2009})}\BibitemShut {NoStop}%
\bibitem [{\citenamefont {Collings}(1980)}]{Sigap}%
  \BibitemOpen
  \bibfield  {author} {\bibinfo {author} {\bibfnamefont {P.~J.}\ \bibnamefont
  {Collings}},\ }\bibfield  {title} {\bibinfo {title} {Simple measurement of
  the band gap in silicon and germanium},\ }\href@noop {} {\bibfield  {journal}
  {\bibinfo  {journal} {American Journal of Physics}\ }\textbf {\bibinfo
  {volume} {48}},\ \bibinfo {pages} {197} (\bibinfo {year} {1980})}\BibitemShut
  {NoStop}%
\bibitem [{\citenamefont {Schinke}\ \emph {et~al.}(2015)\citenamefont
  {Schinke}, \citenamefont {Peest}, \citenamefont {Schmidt}, \citenamefont
  {Brendel}, \citenamefont {Bothe}, \citenamefont {Vogt}, \citenamefont
  {Kr\"{o}ger}, \citenamefont {Winter}, \citenamefont {Schirmacher},
  \citenamefont {Lim}, \citenamefont {Nguyen},\ and\ \citenamefont
  {MacDonald}}]{Sink}%
  \BibitemOpen
  \bibfield  {author} {\bibinfo {author} {\bibfnamefont {C.}~\bibnamefont
  {Schinke}}, \bibinfo {author} {\bibfnamefont {P.~C.}\ \bibnamefont {Peest}},
  \bibinfo {author} {\bibfnamefont {J.}~\bibnamefont {Schmidt}}, \bibinfo
  {author} {\bibfnamefont {R.}~\bibnamefont {Brendel}}, \bibinfo {author}
  {\bibfnamefont {K.}~\bibnamefont {Bothe}}, \bibinfo {author} {\bibfnamefont
  {M.~R.}\ \bibnamefont {Vogt}}, \bibinfo {author} {\bibfnamefont
  {I.}~\bibnamefont {Kr\"{o}ger}}, \bibinfo {author} {\bibfnamefont
  {S.}~\bibnamefont {Winter}}, \bibinfo {author} {\bibfnamefont
  {A.}~\bibnamefont {Schirmacher}}, \bibinfo {author} {\bibfnamefont
  {S.}~\bibnamefont {Lim}}, \bibinfo {author} {\bibfnamefont {H.~T.}\
  \bibnamefont {Nguyen}},\ and\ \bibinfo {author} {\bibfnamefont
  {D.}~\bibnamefont {MacDonald}},\ }\bibfield  {title} {\bibinfo {title}
  {Uncertainty analysis for the coefficient of band-to-band absorption of
  crystalline silicon},\ }\href@noop {} {\bibfield  {journal} {\bibinfo
  {journal} {AIP Advances}\ }\textbf {\bibinfo {volume} {5}},\ \bibinfo {pages}
  {067168} (\bibinfo {year} {2015})}\BibitemShut {NoStop}%
\bibitem [{\citenamefont {Aspnes}\ and\ \citenamefont
  {Studna}(1983)}]{Aspnes1983-gs}%
  \BibitemOpen
  \bibfield  {author} {\bibinfo {author} {\bibfnamefont {D.~E.}\ \bibnamefont
  {Aspnes}}\ and\ \bibinfo {author} {\bibfnamefont {A.~A.}\ \bibnamefont
  {Studna}},\ }\bibfield  {title} {\bibinfo {title} {Dielectric functions and
  optical parameters of si, ge, {GaP}, {GaAs}, {GaSb}, {InP}, {InAs}, and
  {InSb} from 1.5 to 6.0 ev},\ }\href@noop {} {\bibfield  {journal} {\bibinfo
  {journal} {Physical Review B}\ }\textbf {\bibinfo {volume} {27}},\ \bibinfo
  {pages} {985} (\bibinfo {year} {1983})}\BibitemShut {NoStop}%
\bibitem [{\citenamefont {Zhao}\ \emph {et~al.}(1996)\citenamefont {Zhao},
  \citenamefont {Georgakis},\ and\ \citenamefont {Niu}}]{Zhao1996-ua}%
  \BibitemOpen
  \bibfield  {author} {\bibinfo {author} {\bibfnamefont {X.~G.}\ \bibnamefont
  {Zhao}}, \bibinfo {author} {\bibfnamefont {G.~A.}\ \bibnamefont
  {Georgakis}},\ and\ \bibinfo {author} {\bibfnamefont {Q.}~\bibnamefont
  {Niu}},\ }\bibfield  {title} {\bibinfo {title} {Rabi oscillations between
  bloch bands},\ }\href@noop {} {\bibfield  {journal} {\bibinfo  {journal}
  {Physical Review B}\ }\textbf {\bibinfo {volume} {54}},\ \bibinfo {pages}
  {R5235} (\bibinfo {year} {1996})}\BibitemShut {NoStop}%
\bibitem [{\citenamefont {Floss}\ \emph {et~al.}(2018)\citenamefont {Floss},
  \citenamefont {Lemell}, \citenamefont {Wachter}, \citenamefont {Smejkal},
  \citenamefont {Sato}, \citenamefont {Tong}, \citenamefont {Yabana},\ and\
  \citenamefont {Burgd{\"o}rfer}}]{Floss2018-yj}%
  \BibitemOpen
  \bibfield  {author} {\bibinfo {author} {\bibfnamefont {I.}~\bibnamefont
  {Floss}}, \bibinfo {author} {\bibfnamefont {C.}~\bibnamefont {Lemell}},
  \bibinfo {author} {\bibfnamefont {G.}~\bibnamefont {Wachter}}, \bibinfo
  {author} {\bibfnamefont {V.}~\bibnamefont {Smejkal}}, \bibinfo {author}
  {\bibfnamefont {S.~A.}\ \bibnamefont {Sato}}, \bibinfo {author}
  {\bibfnamefont {X.-M.}\ \bibnamefont {Tong}}, \bibinfo {author}
  {\bibfnamefont {K.}~\bibnamefont {Yabana}},\ and\ \bibinfo {author}
  {\bibfnamefont {J.}~\bibnamefont {Burgd{\"o}rfer}},\ }\bibfield  {title}
  {\bibinfo {title} {Ab initio multiscale simulation of high-order harmonic
  generation in solids},\ }\href@noop {} {\bibfield  {journal} {\bibinfo
  {journal} {Physical Review A}\ }\textbf {\bibinfo {volume} {97}},\ \bibinfo
  {pages} {011401} (\bibinfo {year} {2018})}\BibitemShut {NoStop}%
\bibitem [{\citenamefont {Freeman}\ \emph {et~al.}(2022)\citenamefont
  {Freeman}, \citenamefont {Kheifets}, \citenamefont {Yamada}, \citenamefont
  {Yamada},\ and\ \citenamefont {Yabana}}]{Freeman2022-dn}%
  \BibitemOpen
  \bibfield  {author} {\bibinfo {author} {\bibfnamefont {D.}~\bibnamefont
  {Freeman}}, \bibinfo {author} {\bibfnamefont {A.}~\bibnamefont {Kheifets}},
  \bibinfo {author} {\bibfnamefont {S.}~\bibnamefont {Yamada}}, \bibinfo
  {author} {\bibfnamefont {A.}~\bibnamefont {Yamada}},\ and\ \bibinfo {author}
  {\bibfnamefont {K.}~\bibnamefont {Yabana}},\ }\bibfield  {title} {\bibinfo
  {title} {High-order harmonic generation in semiconductors driven at near- and
  mid-infrared wavelengths},\ }\href@noop {} {\bibfield  {journal} {\bibinfo
  {journal} {Physical Review B}\ }\textbf {\bibinfo {volume} {106}},\ \bibinfo
  {pages} {075202} (\bibinfo {year} {2022})}\BibitemShut {NoStop}%
\bibitem [{\citenamefont {Gonze}\ \emph {et~al.}(2020)\citenamefont {Gonze},
  \citenamefont {Amadon}, \citenamefont {Antonius}, \citenamefont {Arnardi},
  \citenamefont {Baguet}, \citenamefont {Beuken}, \citenamefont {Bieder},
  \citenamefont {Bottin}, \citenamefont {Bouchet}, \citenamefont {Bousquet},
  \citenamefont {Brouwer}, \citenamefont {Bruneval}, \citenamefont {Brunin},
  \citenamefont {Cavignac}, \citenamefont {Charraud}, \citenamefont {Chen},
  \citenamefont {C{\^o}t{\'e}}, \citenamefont {Cottenier}, \citenamefont
  {Denier}, \citenamefont {Geneste}, \citenamefont {Ghosez}, \citenamefont
  {Giantomassi}, \citenamefont {Gillet}, \citenamefont {Gingras}, \citenamefont
  {Hamann}, \citenamefont {Hautier}, \citenamefont {He}, \citenamefont
  {Helbig}, \citenamefont {Holzwarth}, \citenamefont {Jia}, \citenamefont
  {Jollet}, \citenamefont {Lafargue-Dit-Hauret}, \citenamefont {Lejaeghere},
  \citenamefont {Marques}, \citenamefont {Martin}, \citenamefont {Martins},
  \citenamefont {Miranda}, \citenamefont {Naccarato}, \citenamefont {Persson},
  \citenamefont {Petretto}, \citenamefont {Planes}, \citenamefont {Pouillon},
  \citenamefont {Prokhorenko}, \citenamefont {Ricci}, \citenamefont
  {Rignanese}, \citenamefont {Romero}, \citenamefont {Schmitt}, \citenamefont
  {Torrent}, \citenamefont {van Setten}, \citenamefont {Van~Troeye},
  \citenamefont {Verstraete}, \citenamefont {Z{\'e}rah},\ and\ \citenamefont
  {Zwanziger}}]{Gonze2020-yq}%
  \BibitemOpen
  \bibfield  {author} {\bibinfo {author} {\bibfnamefont {X.}~\bibnamefont
  {Gonze}}, \bibinfo {author} {\bibfnamefont {B.}~\bibnamefont {Amadon}},
  \bibinfo {author} {\bibfnamefont {G.}~\bibnamefont {Antonius}}, \bibinfo
  {author} {\bibfnamefont {F.}~\bibnamefont {Arnardi}}, \bibinfo {author}
  {\bibfnamefont {L.}~\bibnamefont {Baguet}}, \bibinfo {author} {\bibfnamefont
  {J.-M.}\ \bibnamefont {Beuken}}, \bibinfo {author} {\bibfnamefont
  {J.}~\bibnamefont {Bieder}}, \bibinfo {author} {\bibfnamefont
  {F.}~\bibnamefont {Bottin}}, \bibinfo {author} {\bibfnamefont
  {J.}~\bibnamefont {Bouchet}}, \bibinfo {author} {\bibfnamefont
  {E.}~\bibnamefont {Bousquet}}, \bibinfo {author} {\bibfnamefont
  {N.}~\bibnamefont {Brouwer}}, \bibinfo {author} {\bibfnamefont
  {F.}~\bibnamefont {Bruneval}}, \bibinfo {author} {\bibfnamefont
  {G.}~\bibnamefont {Brunin}}, \bibinfo {author} {\bibfnamefont
  {T.}~\bibnamefont {Cavignac}}, \bibinfo {author} {\bibfnamefont {J.-B.}\
  \bibnamefont {Charraud}}, \bibinfo {author} {\bibfnamefont {W.}~\bibnamefont
  {Chen}}, \bibinfo {author} {\bibfnamefont {M.}~\bibnamefont {C{\^o}t{\'e}}},
  \bibinfo {author} {\bibfnamefont {S.}~\bibnamefont {Cottenier}}, \bibinfo
  {author} {\bibfnamefont {J.}~\bibnamefont {Denier}}, \bibinfo {author}
  {\bibfnamefont {G.}~\bibnamefont {Geneste}}, \bibinfo {author} {\bibfnamefont
  {P.}~\bibnamefont {Ghosez}}, \bibinfo {author} {\bibfnamefont
  {M.}~\bibnamefont {Giantomassi}}, \bibinfo {author} {\bibfnamefont
  {Y.}~\bibnamefont {Gillet}}, \bibinfo {author} {\bibfnamefont
  {O.}~\bibnamefont {Gingras}}, \bibinfo {author} {\bibfnamefont {D.~R.}\
  \bibnamefont {Hamann}}, \bibinfo {author} {\bibfnamefont {G.}~\bibnamefont
  {Hautier}}, \bibinfo {author} {\bibfnamefont {X.}~\bibnamefont {He}},
  \bibinfo {author} {\bibfnamefont {N.}~\bibnamefont {Helbig}}, \bibinfo
  {author} {\bibfnamefont {N.}~\bibnamefont {Holzwarth}}, \bibinfo {author}
  {\bibfnamefont {Y.}~\bibnamefont {Jia}}, \bibinfo {author} {\bibfnamefont
  {F.}~\bibnamefont {Jollet}}, \bibinfo {author} {\bibfnamefont
  {W.}~\bibnamefont {Lafargue-Dit-Hauret}}, \bibinfo {author} {\bibfnamefont
  {K.}~\bibnamefont {Lejaeghere}}, \bibinfo {author} {\bibfnamefont {M.~A.~L.}\
  \bibnamefont {Marques}}, \bibinfo {author} {\bibfnamefont {A.}~\bibnamefont
  {Martin}}, \bibinfo {author} {\bibfnamefont {C.}~\bibnamefont {Martins}},
  \bibinfo {author} {\bibfnamefont {H.~P.~C.}\ \bibnamefont {Miranda}},
  \bibinfo {author} {\bibfnamefont {F.}~\bibnamefont {Naccarato}}, \bibinfo
  {author} {\bibfnamefont {K.}~\bibnamefont {Persson}}, \bibinfo {author}
  {\bibfnamefont {G.}~\bibnamefont {Petretto}}, \bibinfo {author}
  {\bibfnamefont {V.}~\bibnamefont {Planes}}, \bibinfo {author} {\bibfnamefont
  {Y.}~\bibnamefont {Pouillon}}, \bibinfo {author} {\bibfnamefont
  {S.}~\bibnamefont {Prokhorenko}}, \bibinfo {author} {\bibfnamefont
  {F.}~\bibnamefont {Ricci}}, \bibinfo {author} {\bibfnamefont {G.-M.}\
  \bibnamefont {Rignanese}}, \bibinfo {author} {\bibfnamefont {A.~H.}\
  \bibnamefont {Romero}}, \bibinfo {author} {\bibfnamefont {M.~M.}\
  \bibnamefont {Schmitt}}, \bibinfo {author} {\bibfnamefont {M.}~\bibnamefont
  {Torrent}}, \bibinfo {author} {\bibfnamefont {M.~J.}\ \bibnamefont {van
  Setten}}, \bibinfo {author} {\bibfnamefont {B.}~\bibnamefont {Van~Troeye}},
  \bibinfo {author} {\bibfnamefont {M.~J.}\ \bibnamefont {Verstraete}},
  \bibinfo {author} {\bibfnamefont {G.}~\bibnamefont {Z{\'e}rah}},\ and\
  \bibinfo {author} {\bibfnamefont {J.~W.}\ \bibnamefont {Zwanziger}},\
  }\bibfield  {title} {\bibinfo {title} {The abinitproject: Impact, environment
  and recent developments},\ }\href@noop {} {\bibfield  {journal} {\bibinfo
  {journal} {Computer Physics Communications}\ }\textbf {\bibinfo {volume}
  {248}},\ \bibinfo {pages} {107042} (\bibinfo {year} {2020})}\BibitemShut
  {NoStop}%
\bibitem [{\citenamefont {Romero}\ \emph {et~al.}(2020)\citenamefont {Romero},
  \citenamefont {Allan}, \citenamefont {Amadon}, \citenamefont {Antonius},
  \citenamefont {Applencourt}, \citenamefont {Baguet}, \citenamefont {Bieder},
  \citenamefont {Bottin}, \citenamefont {Bouchet}, \citenamefont {Bousquet},
  \citenamefont {Bruneval}, \citenamefont {Brunin}, \citenamefont {Caliste},
  \citenamefont {C{\^o}t{\'e}}, \citenamefont {Denier}, \citenamefont {Dreyer},
  \citenamefont {Ghosez}, \citenamefont {Giantomassi}, \citenamefont {Gillet},
  \citenamefont {Gingras}, \citenamefont {Hamann}, \citenamefont {Hautier},
  \citenamefont {Jollet}, \citenamefont {Jomard}, \citenamefont {Martin},
  \citenamefont {Miranda}, \citenamefont {Naccarato}, \citenamefont {Petretto},
  \citenamefont {Pike}, \citenamefont {Planes}, \citenamefont {Prokhorenko},
  \citenamefont {Rangel}, \citenamefont {Ricci}, \citenamefont {Rignanese},
  \citenamefont {Royo}, \citenamefont {Stengel}, \citenamefont {Torrent},
  \citenamefont {van Setten}, \citenamefont {Van~Troeye}, \citenamefont
  {Verstraete}, \citenamefont {Wiktor}, \citenamefont {Zwanziger},\ and\
  \citenamefont {Gonze}}]{Romero2020-hq}%
  \BibitemOpen
  \bibfield  {author} {\bibinfo {author} {\bibfnamefont {A.~H.}\ \bibnamefont
  {Romero}}, \bibinfo {author} {\bibfnamefont {D.~C.}\ \bibnamefont {Allan}},
  \bibinfo {author} {\bibfnamefont {B.}~\bibnamefont {Amadon}}, \bibinfo
  {author} {\bibfnamefont {G.}~\bibnamefont {Antonius}}, \bibinfo {author}
  {\bibfnamefont {T.}~\bibnamefont {Applencourt}}, \bibinfo {author}
  {\bibfnamefont {L.}~\bibnamefont {Baguet}}, \bibinfo {author} {\bibfnamefont
  {J.}~\bibnamefont {Bieder}}, \bibinfo {author} {\bibfnamefont
  {F.}~\bibnamefont {Bottin}}, \bibinfo {author} {\bibfnamefont
  {J.}~\bibnamefont {Bouchet}}, \bibinfo {author} {\bibfnamefont
  {E.}~\bibnamefont {Bousquet}}, \bibinfo {author} {\bibfnamefont
  {F.}~\bibnamefont {Bruneval}}, \bibinfo {author} {\bibfnamefont
  {G.}~\bibnamefont {Brunin}}, \bibinfo {author} {\bibfnamefont
  {D.}~\bibnamefont {Caliste}}, \bibinfo {author} {\bibfnamefont
  {M.}~\bibnamefont {C{\^o}t{\'e}}}, \bibinfo {author} {\bibfnamefont
  {J.}~\bibnamefont {Denier}}, \bibinfo {author} {\bibfnamefont
  {C.}~\bibnamefont {Dreyer}}, \bibinfo {author} {\bibfnamefont
  {P.}~\bibnamefont {Ghosez}}, \bibinfo {author} {\bibfnamefont
  {M.}~\bibnamefont {Giantomassi}}, \bibinfo {author} {\bibfnamefont
  {Y.}~\bibnamefont {Gillet}}, \bibinfo {author} {\bibfnamefont
  {O.}~\bibnamefont {Gingras}}, \bibinfo {author} {\bibfnamefont {D.~R.}\
  \bibnamefont {Hamann}}, \bibinfo {author} {\bibfnamefont {G.}~\bibnamefont
  {Hautier}}, \bibinfo {author} {\bibfnamefont {F.}~\bibnamefont {Jollet}},
  \bibinfo {author} {\bibfnamefont {G.}~\bibnamefont {Jomard}}, \bibinfo
  {author} {\bibfnamefont {A.}~\bibnamefont {Martin}}, \bibinfo {author}
  {\bibfnamefont {H.~P.~C.}\ \bibnamefont {Miranda}}, \bibinfo {author}
  {\bibfnamefont {F.}~\bibnamefont {Naccarato}}, \bibinfo {author}
  {\bibfnamefont {G.}~\bibnamefont {Petretto}}, \bibinfo {author}
  {\bibfnamefont {N.~A.}\ \bibnamefont {Pike}}, \bibinfo {author}
  {\bibfnamefont {V.}~\bibnamefont {Planes}}, \bibinfo {author} {\bibfnamefont
  {S.}~\bibnamefont {Prokhorenko}}, \bibinfo {author} {\bibfnamefont
  {T.}~\bibnamefont {Rangel}}, \bibinfo {author} {\bibfnamefont
  {F.}~\bibnamefont {Ricci}}, \bibinfo {author} {\bibfnamefont {G.-M.}\
  \bibnamefont {Rignanese}}, \bibinfo {author} {\bibfnamefont {M.}~\bibnamefont
  {Royo}}, \bibinfo {author} {\bibfnamefont {M.}~\bibnamefont {Stengel}},
  \bibinfo {author} {\bibfnamefont {M.}~\bibnamefont {Torrent}}, \bibinfo
  {author} {\bibfnamefont {M.~J.}\ \bibnamefont {van Setten}}, \bibinfo
  {author} {\bibfnamefont {B.}~\bibnamefont {Van~Troeye}}, \bibinfo {author}
  {\bibfnamefont {M.~J.}\ \bibnamefont {Verstraete}}, \bibinfo {author}
  {\bibfnamefont {J.}~\bibnamefont {Wiktor}}, \bibinfo {author} {\bibfnamefont
  {J.~W.}\ \bibnamefont {Zwanziger}},\ and\ \bibinfo {author} {\bibfnamefont
  {X.}~\bibnamefont {Gonze}},\ }\bibfield  {title} {\bibinfo {title} {{ABINIT}:
  Overview and focus on selected capabilities},\ }\href@noop {} {\bibfield
  {journal} {\bibinfo  {journal} {Jounal of Chemical Physics}\ }\textbf
  {\bibinfo {volume} {152}},\ \bibinfo {pages} {124102} (\bibinfo {year}
  {2020})}\BibitemShut {NoStop}%
\bibitem [{\citenamefont {Kokalj}(2003)}]{Kokalj2003-nf}%
  \BibitemOpen
  \bibfield  {author} {\bibinfo {author} {\bibfnamefont {A.}~\bibnamefont
  {Kokalj}},\ }\bibfield  {title} {\bibinfo {title} {Computer graphics and
  graphical user interfaces as tools in imulations of matter at the atomic
  scale},\ }\href@noop {} {\bibfield  {journal} {\bibinfo  {journal}
  {Computational Materials Science}\ }\textbf {\bibinfo {volume} {28}},\
  \bibinfo {pages} {155} (\bibinfo {year} {2003})}\BibitemShut {NoStop}%
\end{thebibliography}
%


\end{document}